\documentclass{article}

\usepackage{PRIMEarxiv}

\usepackage[utf8]{inputenc} 
\usepackage[T1]{fontenc}    
\usepackage{hyperref}       
\usepackage{url}            
\usepackage{booktabs}       
\usepackage{amsfonts}       
\usepackage{nicefrac}       
\usepackage{lipsum}
\usepackage{fancyhdr}       
\usepackage{graphicx}       
\graphicspath{{media/}}     
\usepackage[square,numbers]{natbib}
\usepackage{tabularx}  
\usepackage{changepage}
\usepackage{tcolorbox}
\usepackage{listings}
\usepackage{lscape}  
\lstdefinelanguage{Solidity}{
    keywords={contract, import, pragma, function, return, view, public, external, pure, uint, uint256, address, payable, returns, this, selfdestruct, require, if, else, for, emit, event},
    keywordstyle=\color{blue},
    ndkeywords={},
    ndkeywordstyle=\color{darkgray},
    identifierstyle=\color{black},
    sensitive=false,
    comment=[l]{//},
    morecomment=[s]{/*}{*/},
    commentstyle=\color{olive},
    stringstyle=\color{red},
    morestring=[b]',
    morestring=[b]",
    showstringspaces=false
}
\usepackage{setspace}
\usepackage{longtable}
\usepackage{csvsimple}
\usepackage[nopatch=eqnum]{microtype}

\title{Vulnerability Detection in Smart Contracts: A Comprehensive Survey

}

\author{
    Christopher De Baets \\
    University of Sydney \\
    Sydney Australia\\
    \texttt{chde0116@uni.sydney.edu.au}
    \And
    Basem Suleiman \\
    University of New South Wales Sydney \\
    Sydney, Australia\\
    \texttt{b.suleiman@unsw.edu.au}
    \And
    Armin Chitizadeh \\
    University of New South Wales Sydney \\
    Sydney, Australia \\
    \texttt{a.chitizadeh@unsw.edu.au} \\
    \And
    Imran Razzak \\
    University of New South Wales Sydney \\
    Sydney, Australia\\
    \texttt{imran.razzak@unsw.edu.au}
}

\begin{document}
\maketitle

\begin{abstract}
In the growing field of blockchain technology, smart contracts exist as transformative digital agreements that execute transactions autonomously in decentralised networks. However, these contracts face challenges in the form of security vulnerabilities, posing significant financial and operational risks. While traditional methods to detect and mitigate vulnerabilities in smart contracts are limited due to a lack of comprehensiveness and effectiveness, integrating advanced machine learning technologies presents an attractive approach to increasing effective vulnerability countermeasures. We endeavour to fill an important gap in the existing literature by conducting a rigorous systematic review, exploring the intersection between machine learning and smart contracts. Specifically, the study examines the potential of machine learning techniques to improve the detection and mitigation of vulnerabilities in smart contracts. 
We analysed 88 articles published between 2018 and 2023 from the following databases: IEEE, ACM, ScienceDirect, Scopus, and Google Scholar. The findings reveal that classical machine learning techniques, including KNN, RF, DT, XG-Boost, and SVM, outperform static tools in vulnerability detection. Moreover, multi-model approaches integrating deep learning and classical machine learning show significant improvements in precision and recall, while hybrid models employing various techniques achieve near-perfect performance in vulnerability detection accuracy.

By integrating state-of-the-art solutions, this work synthesises current methods, thoroughly investigates research gaps, and suggests directions for future studies. The insights gathered from this study are intended to serve as a seminal reference for academics, industry experts, and bodies interested in leveraging machine learning to enhance smart contract security.
\end{abstract}



\section{Introduction}

Conceived as an advanced iteration of the rudimentary blockchain technology that underpins Bitcoin, Ethereum, conceptualised by Vitalik Buterin, serves as a decentralised platform engineered explicitly for the formulation and execution of smart contracts - self-executing digital agreements with terms impressed in code \cite{buterin2013ethereum}. This innovation moves Ethereum beyond the confines of Bitcoin's limited scripting capabilities, offering a framework conducive to a diverse spectrum of decentralised applications (DApps). Stemming from its Turing-complete architecture, explored in Gavin Wood's seminal paper \cite{wood2014ethereum}, Ethereum's computational flexibility can theoretically tackle any problem, given sufficient resources. Unlike Bitcoin, which serves primarily as a digital currency, Ethereum extends the practical utility of blockchain technology, facilitating a network conducive to decentralised services. At the core of this functionality lies the Ethereum Virtual Machine (EVM), responsible for the autonomous execution of smart contracts and removing the necessity for centralised intermediaries. Ethereum's versatile platform has catalysed advancements across many sectors, including finance and autonomous governance \cite{luu2016making}. Given Ethereum's versatility and the imperative of securing transactions, a thorough assessment of how the blockchain's core mechanisms can be fortified is warranted, steering our examination towards the pivotal role of smart contracts and how their security may be strengthened.


As blockchain adoption proliferates, the security of users' funds becomes crucial in order to uphold the trust of users. Given the importance of identifying ways in which different technologies can enhance the capabilities and address the limitations of systems, a systematic literature review in this particular niche is needed. While there exist some systematic literature review, however, they are either too old to be relevant to current needs or they did not focus on enough range of machine learning models. The goal of this article is to consolidate existing knowledge, identify research gaps, and provide a granular understanding of the current state of research on machine learning-applied smart contract vulnerability detection. This will help guide future research and development in this field, and create a comprehensive research repository for stakeholders in blockchain security.



In light of the motivations of this research, we have formulated the following research questions to guide our investigation:

\begin{enumerate}

    \item[(RQ1)] What are the state-of-the-art machine learning techniques for identifying and mitigating specific vulnerabilities in smart contracts?
    \item[(RQ2)] What machine learning algorithms have been applied to smart contract vulnerability detection, and how do they differ in efficacy and limitations when detecting vulnerabilities?
    \item[(RQ3)] What are the current research gaps and future work opportunities in the application of machine learning for enhancing the detection and mitigation of vulnerabilities in smart contracts?
    
\end{enumerate}


In summary, the contributions of this work are threefold:

\begin{itemize}
    \item Developed a comprehensive repository of research papers on machine learning for smart contract vulnerability detection, manifesting in a practical resource for parties involved in blockchain security.
    \item Analysed and synthesised current machine learning models targeting smart contract vulnerabilities and demonstrated how machine learning can improve the detection and mitigation of vulnerabilities.
    \item Identified research gaps, guiding further research and future directions in smart contract security.
\end{itemize}


The remaining structure of this article is as follows. The related work section (Section \ref{sec:relatedwork}) discusses the use of machine learning in detecting vulnerabilities in smart contracts, highlighting gaps in previous literature reviews and underscoring the comprehensive approach of the current study. In the background section (Section \ref{sec:background}), we establish foundational knowledge regarding Ethereum, smart contract vulnerabilities, and existing traditional detection methods. We also introduce several standard machine learning techniques, including traditional models and neural networks. In the methodology section (Section \ref{sec:methodology}),  we formulate research questions, outline the research methodology, describing our search strategy, databases, search queries and criteria pertinent to the systematic review process. We also detail our quality assessment considerations, data extraction, paper categorisation and data synthesis, bearing in mind the limitations of our study. In the results section (Section \ref{sec:results}), we detail the findings of our systematic literature review, including the source collection process, publication distribution over time, and categorization of machine learning models to highlight common vulnerability detection approaches and explore popular vulnerabilities. It follows by  the synthesis of results section (Section \ref{sec:synResults}) in which we dive deeper into the papers retrieved from our search, after which we segue into the discussion section (Section \ref{chap:discussion}) to critically analyse the current corpus of research, examining the diverse use cases of machine learning techniques in enhancing smart contract vulnerability detection. The article concludes in section \ref{sec:conclusion}, in which we summarise our contributions.

\section{Related Work}\label{sec:relatedwork}
The use of machine learning approaches in smart contract vulnerability detection has gained attention in recent years. However, previous literature review papers on this topic have either covered only a few articles on machine learning approaches or completely missed them.

\cite{singh2020blockchain} conducted a rigorous systematic review of 35 research papers from 2015 to 2019 and discovered the most commonly used technique was theorem proving. None of the reviewed papers considered any machine learning approaches. \cite{vacca2021systematic} reviewed only two machine learning related papers, one from 2018 and other from 2019. \cite{piantadosi2023detecting} conducted a systematic literature review with a section on machine learning-based methods, but only considered six machine learning approaches, five from 2019 and one from 2020. Similarly, \cite{vacca2021systematic} reviewed only two papers with machine-learning approaches, one from 2018 and other from 2019. \cite{chu2023survey} explored smart contract security from multiple perspective including vulnerability detection. However, out of 49 articles that they covered, only 21 used a machine learning approach, which is much smaller than the 88 articles reviewed in this paper. This paper has the advantage of covering a more recent body of research, as it has discovered many papers published in 2021 or later on this rapidly evolving topic.

Some review articles have focuses on tools on detecting vulnerabilities rather than algorithms. Examples include \cite{kushwaha2022systematic, kushwaha2022ethereum,zhou2022state} and \cite{di2024evolution}. Most vulnerability detection tools employ static methods. Even if they utilise machine learning within their tool, the approach is still less flexible and customisable. As a result, the returned results may not be as good as using a dedicated machine learning approach.

More recent reviews have examined machine learning approaches; however, they are not comprehensive in their scope and do not offer a complete overview of the existing literature. \cite{surucu2022survey} reviewed several Machine Learning approaches in smart contract vulnerability detection and sets the ground for further development in this field. However, it is not a systematic literature review and lacks focused and comprehensiveness. This difference highlights a unique aspect of our research, emphasizing the distinct contribution our work makes.\cite{Jiang2023} provides an exploration of machine learning applications in detecting smart contract vulnerabilities. They categorise common vulnerabilities associated with smart contracts, and delve into the formalised tools currently available for vulnerability detection. They also highlight the strengths and limitations of machine learning methods concerning their application in smart contract security. This paper intersects significantly with our scope in that it discusses many similar themes; however, this survey is not exhaustive in its approach and does not provide a full account of the published literature. This gap offers a distinct degree of differentiation between our studies, reinforcing the unique contribution of our work. Another paper contributing to the discourse in machine learning applications to smart contract security is \cite{Krichen2023}, which provides an overview of smart contracts and the array of attacks blockchain infrastructures are prone to. A significant portion of their analysis is dedicated to how artificial intelligence strategies can mitigate various attacks. This paper also aligns closely with our themes. However, again, despite its insights, the paper does not provide an exhaustive review of the work within this domain, highlighting the distinctiveness of our study.

\section{Background} \label{sec:background}
This section presents contextual and prerequisite information critical for understanding our work. We first discuss the rudimentary premise of smart contracts, their inherent vulnerabilities and existing vulnerability detection tools. Acknowledging the limitations of these tools, we then segue into a basic introduction of the standard classical machine learning and neural network models. Collectively, this briefing establishes a foundational understanding of the concepts elaborated upon in our review.

\subsection{Smart Contracts}

In contrast to conventional contracts, which predominantly exist in textual and oral formats and are contingent on third-party oracles, Ethereum's smart contracts are self-executing and self-regulating by invention. Once instantiated on the blockchain, smart contracts autonomously activate when predefined conditions are met, thereby minimising transaction costs instigated by third parties and substantially reducing the risk of fraudulent activities and human error. A smart contract on the Ethereum platform is structured around three foundational elements: the parties interfacing directly with the contract, who initiate and stipulate certain conditions; the object of the agreement, delineating whatever assets, criteria or services must be met by the parties involved; and the explicit terms and conditions, which are codified into executable logic \cite{wang_overview_2018}. The development of the actual smart contract begins with stakeholders negotiating to define its constraints and obligations. Once a consensus has been reached, legal professionals will draft a provisional document, which software engineers convert into machine-readable code. This conversion comprises iterative conceptual design, coding, and validation stages, necessitating collaboration from diverse entities. Once the smart contract has been vetted, it is enshrined onto the Ethereum blockchain, rendering it immutable and transparent to all relevant parties. The digitised assets of the parties are placed in escrow through the suspension of their respective digital wallets whilst the contract undergoes execution. In this stage, its conditions are constantly checked. As specific conditions are met, functions in the contract are executed, and transactions are authenticated by network miners and are recorded on the blockchain. Once all contractual terms are met, the transfer of digital assets is processed, and the wallets are unfrozen, marking the completion of the contract's lifecycle \cite{zheng_overview_2020}.

In Ethereum's ecosystem, Solidity is the language of choice for coding smart contracts, given its extensive functionalities and robust development environment catered toward decentralised applications. However, the properties that enable a smart contract's flexibility also bestow complexities that render them susceptible to vulnerabilities and coding errors \cite{atzei2017survey}. This inherent duality between flexibility and security underscores the imperative to systematically investigate how these susceptibilities may be mitigated.

\subsection{Smart Contract Vulnerabilities}

This part will consolidate our understanding of smart contract vulnerabilities by covering three essential areas. We will first identify the most common types of vulnerabilities, examine how they emerge, and evaluate how they can affect these contracts' integrity and reliability. Then, we will review existing methods for detecting these vulnerabilities, delineating their limitations and effectiveness. Finally, we will discuss the potential dangers of leaving such vulnerabilities unaddressed, highlighting the need for more practical and innovative solutions, such as those pertaining to new machine learning techniques. Examining these vulnerabilities in detail is a critical prerequisite for the discussion of using machine learning techniques to address these challenges effectively.

\subsubsection{Reentrancy}
\begin{figure}[h!]
    \begin{tcolorbox}[colback=white, colframe=black, boxrule=1pt, boxsep=5pt, left=5pt, right=5pt, top=5pt, bottom=5pt]
    
        \begin{lstlisting}[language=Solidity, basicstyle=\ttfamily\scriptsize, lineskip=0.25mm]]
        pragma solidity ^0.8.0;
        contract Vulnerable {
            mapping(address => uint) public balances;
        
            function deposit() public payable {
                balances[msg.sender] += msg.value;
            }
        
            function withdraw(uint _amount) public {
                require(balances[msg.sender] >= _amount);
                
                /*
                    Balance is not updated until after the external call.
                    The attacker can constantly enter the function and deplete 
                    the contract balance before the initial call can be completed.
                */                
                (bool sent, ) = msg.sender.call{value: _amount}("");
                require(sent, "Could not send Ether");
                balances[msg.sender] -= _amount;
                emit Transfer(msg.sender, _amount);
            }
        }    
        \end{lstlisting}
    \end{tcolorbox}
    \caption{Reentrancy vulnerability.}
    \label{fig:v_reentrancy}
\end{figure}

Figure \ref{fig:v_reentrancy} demonstrates an example of a reentrancy vulnerability. In this contract, the \textit{withdraw} function contains a flaw in that it performs an external call to \textit{msg.sender} before updating the \textit{balances} array. Allowing such a contract to be deployed could result in an attacker exploiting this flaw by withdrawing more ether than intended, depleting the balance in the contract. 

\subsubsection{Integer Underflow and Overflow (Arithmetic)}
\begin{figure}[h!]
    \begin{tcolorbox}[colback=white, colframe=black, boxrule=1pt, boxsep=5pt, left=5pt, right=5pt, top=5pt, bottom=5pt]
        \begin{lstlisting}[language=Solidity, basicstyle=\ttfamily\scriptsize, lineskip=0.25mm]]
        pragma solidity ^0.6.0;
        contract Vulnerable {
            uint public value;
        
            function unsafeUpdate(uint _add, uint _subtract) public {      
            
                /*
                    Integer overflow: if _add is big enough, 
                    the value will wrap around.
                */
                value += _add;

                /*
                    Integer underflow: if _subtract is greater than value, 
                    the value will wrap around.
                */
                value -= _subtract;
            }
        }    
        \end{lstlisting}
    \end{tcolorbox}
    \caption{Integer overflow and underflow vulnerability.}
    \label{fig:v_int_overflow_underflow}
\end{figure}

The contract presented in figure  \ref{fig:v_int_overflow_underflow} presents two arithmetic operation vulnerabilities: integer overflow and underflow. The \textit{unsafeUpdate} function has no safeguard to prevent an overflow or underflow error from occurring, depending on the value of \textit{\_add} and \textit{\_subtract}. If \textit{\_add} is sufficiently large, \textit{value} will wrap and reset to a smaller number due to the capacity limitations of the \textit{uint} data type. The same goes for \textit{value} if \textit{\_subtract} is larger than \textit{value}. This is considered a significant security risk, as it can transform the state of the contract in unexpected ways.

\subsubsection{Timestamp Dependency}
\begin{figure}[h!]
    \begin{tcolorbox}[colback=white, colframe=black, boxrule=1pt, boxsep=5pt, left=5pt, right=5pt, top=5pt, bottom=5pt]
        \begin{lstlisting}[language=Solidity, basicstyle=\ttfamily\scriptsize, lineskip=0.25mm]]
        pragma solidity ^0.8.0;
        contract Vulnerable {
            uint public prevClaimTime;
    
            // Vulnerable: relies on block.timestamp.
            function claim() public {
                require(block.timestamp > prevClaimTime + 1, "Unable to claim yet");
                prevClaimTime = block.timestamp;
            }
        }    
        \end{lstlisting}
    \end{tcolorbox}
    \caption{Timestamp dependency vulnerability.}
    \label{fig:v_timestamp_dependency}
\end{figure}

The code depicted in \ref{fig:v_timestamp_dependency} illustrates a timestamp dependency vulnerability, in which the \textit{block.timestamp} variable is used to determine whether a user can execute the \textit{claim} function, imposing a time constraint with the \textit{require} keyword. The problem with this is that miners hold a degree of leeway when manipulating the block timestamp, rendering the logic of this contract precarious. Theoretically, if this contract were deployed, a miner could adjust the timestamp to circumvent the \textit{require} condition to interfere with or exploit other parts of the contract.

\subsubsection{tx.origin Dependence}
\begin{figure}[h!]
    \begin{tcolorbox}[colback=white, colframe=black, boxrule=1pt, boxsep=5pt, left=5pt, right=5pt, top=5pt, bottom=5pt]
        \begin{lstlisting}[language=Solidity, basicstyle=\ttfamily\scriptsize, lineskip=0.25mm]]
        pragma solidity ^0.8.0;
        contract Vulnerable {
            address public owner;
            mapping(address => uint) public balances;
    
            constructor() public {
                owner = msg.sender;
            }
    
            // Vulnerable function: relies on tx.origin for authorisation.
            function unsafeTransfer(address _to, uint _amount) public {
                require(tx.origin == owner, "Not authorised");
                balances[_to] += _amount;
            }
        }    
        \end{lstlisting}
    \end{tcolorbox}
    \caption{tx.origin Dependence vulnerability.}
    \label{fig:v_txorigin_dependency}
\end{figure}

Figure \ref{fig:v_txorigin_dependency} presents a tx.origin dependence vulnerability, in which the \textit{unsafeTransfer} function depends on \textit{tx.origin} for authorisation checks, as opposed to \textit{msg.sender}. The difference between the two is that \textit{tx.origin} represents the creator of the transaction chain, whereas \textit{msg.sender} represents the immediate sender of the transaction. If an attacker deceived the owner into making a transaction that calls an intermediary contract, the attacker's contract could then call \textit{unsafeTransfer} because \textit{tx.origin} would still be the owner's address.

\subsubsection{Unchecked Low Level Call}
\begin{figure}[h!]
    \begin{tcolorbox}[colback=white, colframe=black, boxrule=1pt, boxsep=5pt, left=5pt, right=5pt, top=5pt, bottom=5pt]    
        \begin{lstlisting}[language=Solidity, basicstyle=\ttfamily\scriptsize, lineskip=0.25mm]]
        pragma solidity ^0.8.0;
        contract Vulnerable {
            mapping(address => uint) public balances;
    
            // Vulnerable function: doesn't check result of call.
            function unsafeWithdraw(uint _amount) public {
                require(balances[msg.sender] >= _amount, "Insufficient funds");
    
                // Call without a check for success or failure.
                msg.sender.call{value: _amount}("");
                balances[msg.sender] -= _amount;
            }
        }    
        \end{lstlisting}
    \end{tcolorbox}
    \caption{Unchecked Low Level Call vulnerability.}
    \label{fig:v_unchecked_call}
\end{figure}

An unchecked low level call vulnerability resides in the code of figure \ref{fig:v_unchecked_call}. The \textit{unsafeWithdraw} function executes an external call to \textit{msg.sender} without checking the outcome of the call. The \textit{.call()} method returns a boolean value, which should be checked to inform any subsequent logic. In this case, because the contract does not check the result, the \textit{\_amount} is deducted from \textit{balances} irrespective of the outcome of the \textit{.call()} method. This could result in inconsistencies in the state and renders it susceptible to monetary inaccuracies.

\subsubsection{Front Running}
\begin{figure}[h!]
    \begin{tcolorbox}[colback=white, colframe=black, boxrule=1pt, boxsep=5pt, left=5pt, right=5pt, top=5pt, bottom=5pt]
        \begin{lstlisting}[language=Solidity, basicstyle=\ttfamily\scriptsize, lineskip=0.25mm]]
        pragma solidity ^0.8.0;
        contract Vulnerable {
            uint public price;
    
            // Vulnerable function: susceptible to front running.
            function setPrice(uint _newPrice) public {
                price = _newPrice;
            }
        }    
        \end{lstlisting}
    \end{tcolorbox}
    \caption{Front running vulnerability.}
    \label{fig:v_front_running}
\end{figure}

In figure \ref{fig:v_front_running}, \textit{setPrice} is designed to update the \textit{price} variable; however, it is vulnerable to front running due to the transparency of transactions and the blockchain architecture. As a user updates the \textit{price} variable by sending a transaction, that transaction is stored temporarily in the Ethereum transaction pool before it gets mined into a block. In this state of limbo, attackers can execute their own transaction to update \textit{price} with a higher gas fee. Ethereum prioritises transactions offering higher gas fees, so the attacker's transaction will likely be confirmed first, updating the \textit{price} of the initial transaction. As with standard front running attacks, this vulnerability can result in exploitation and manipulation, as the attacker has the opportunity to benefit from the price change before others have the time to react.

\subsubsection{Access Control}
\begin{figure}[h!]
    \begin{tcolorbox}[colback=white, colframe=black, boxrule=1pt, boxsep=5pt, left=5pt, right=5pt, top=5pt, bottom=5pt]

        \begin{lstlisting}[language=Solidity, basicstyle=\ttfamily\scriptsize, lineskip=0.25mm]]
        pragma solidity ^0.8.0;
        contract Vulnerable {
            uint public price;
            address public owner;
    
            // Vulnerable function: incomplete access control.
            function setPrice(uint _newPrice) public {
                require(msg.sender == owner , "Unauthorised");
                price = _newPrice;
            }
        }    
        \end{lstlisting}
    \end{tcolorbox}
    \caption{Access control vulnerability.}
    \label{fig:v_access_control}
\end{figure}

In figure \ref{fig:v_access_control}, an authorisation mechanism appears to be implemented through the \textit{require} statement in the \textit{setPrice} function; however, the contract showcases an access control vulnerability. Although the \textit{setPrice} function aims to restrict access to \textit{owner} by comparing it with \textit{msg.sender}, the contract is missing a constructor to initialise the \textit{owner} address, resulting in \textit{owner} being assigned \textit{address(0)}. As a result, the function is locked to all users and can only be granted permission by someone sending a transaction from \textit{address(0)}.

\subsubsection{Denial of Service}
\begin{figure}[h!]
    \begin{tcolorbox}[colback=white, colframe=black, boxrule=1pt, boxsep=5pt, left=5pt, right=5pt, top=5pt, bottom=5pt]
        \begin{lstlisting}[language=Solidity, basicstyle=\ttfamily\scriptsize, lineskip=0.25mm]]
        pragma solidity ^0.8.0;
        contract Vulnerable {
            address[] public users;
    
            function addUser(address user) public {
                users.push(user)
            }

            /*
                Vulnerable function: an attacker could populate 'users' array
                with many entries, making the 'withdraw' function
                computationally expensive.
            */            
            function withdraw() public {
                for (uint i = 0; i < users.length; i++) {
                    // Logic...
                }
            }
        }    
        \end{lstlisting}
    \end{tcolorbox}
    \caption{DoS vulnerability.}
    \label{fig:v_dos}
\end{figure}

The \textit{withdraw} function in figure \ref{fig:v_dos} utilises a loop that iterates over an array of users. Although this is a seemingly innocuous operation, such a design pattern produces an attack vector, whereby an attacker may exploit the \textit{addUser} function to store a massive number of entries, making the \textit{withdraw} function too computationally expensive to be considered a feasible call, effectively blocking users from calling the \textit{withdraw} function successfully, disrupting its indeed functionality.

\subsubsection{Infinite Loop}

\begin{figure}[h!]
    \begin{tcolorbox}[colback=white, colframe=black, boxrule=1pt, boxsep=5pt, left=5pt, right=5pt, top=5pt, bottom=5pt]
        \begin{lstlisting}[language=Solidity, basicstyle=\ttfamily\scriptsize, lineskip=0.25mm]]
        pragma solidity ^0.8.0;
        contract Vulnerable {
            uint public count;
    
            // Vulnerable function: infinite loop.
            function increment() public {
                while(true) {
                    counter += 1;
                }
            }
        }    
        \end{lstlisting}
    \end{tcolorbox}
    \caption{Infinite loop vulnerability.}
    \label{fig:v_infinite_loop}
\end{figure}

The \textit{increment} function in \ref{fig:v_infinite_loop} contains a \textit{while} loop without a proper termination condition. The infinite loop vulnerability can result in the exhaustion of all the gas allotted to a transaction without successfully completing the function call, rendering the contract completely dysfunctional for any practical application.
 
\subsubsection{Self Destruct}

\begin{figure}[h!]
    \begin{tcolorbox}[colback=white, colframe=black, boxrule=1pt, boxsep=5pt, left=5pt, right=5pt, top=5pt, bottom=5pt]
        \begin{lstlisting}[language=Solidity, basicstyle=\ttfamily\scriptsize, lineskip=0.25mm]]
        pragma solidity ^0.8.0;
        contract Vulnerable {
            address public owner;
    
            constructor() {
                owner = msg.sender;
            }
    
            // Vulnerable function: self destruct is triggerable by owner.
            function kill() public {
                require(msg.sender == owner, "Unauthorised");
                selfdestruct(payable(owner));
            }
        }    
        \end{lstlisting}
    \end{tcolorbox}
    \caption{Self destruct vulnerability.}
    \label{fig:v_self_destruct}
\end{figure}

The \textit{selfdestruct} method in the \textit{kill} function in figure \ref{fig:v_self_destruct} enables the \textit{owner} to terminate the contract irrevocably and sends the remaining Ether to the \textit{owner} address. Although this call has legitimate purposes, there are risks associated in that once the call has been invoked, the contract becomes completely inoperable and is removed from the blockchain, which could result in a loss of data and critical functionality in other smart contracts. A malicious individual could also exploit such a feature if they were to gain access to the \textit{owner} account.

\subsubsection{Delegate Call}

\begin{figure}[h!]
    \begin{tcolorbox}[colback=white, colframe=black, boxrule=1pt, boxsep=5pt, left=5pt, right=5pt, top=5pt, bottom=5pt]
        \begin{lstlisting}[language=Solidity, basicstyle=\ttfamily\scriptsize, lineskip=0.25mm, keepspaces=true]
        pragma solidity ^0.8.0;
        contract Vulnerable {
            address public owner;
    
            constructor() {
                owner = msg.sender;
            }
    
            // Vulnerable function: delegate call to external contracts.
            function execute(address _target, bytes calldata _data) public {
                require(msg.sender == owner, "Unauthorised");
                (bool success, ) = _target.delegatecall(_data);
                require(success, "Delegate call failed");
            }
        }    
        \end{lstlisting}
    \end{tcolorbox}
    \caption{Delegate call vulnerability.}
    \label{fig:v_delegate_call}
\end{figure}

The \textit{execute} function in figure \ref{fig:v_delegate_call} allows the \textit{owner} to execute arbitrary code on the contract's behalf by calling the \textit{delegatecall} method on \textit{\_target}. Although this feature enables a level of flexibility, it carries several risks. \textit{delegatecall} enables the code of the \textit{\_target} contract to access and modify any state variables of the calling contract. If the external contract is maliciously designed, the state of the contract becomes compromised.

\subsection{Existing Vulnerability Detection Tools} \label{sec:existing_tools}

\cite{durieux_empirical_2020} performs an empirical review of traditional vulnerability analysis techniques, featuring a critical evaluation of widely recognised tools such as Oyente \cite{oyente}, Mythril \cite{mythril}, Securify \cite{securify} and SmartCheck \cite{smartcheck}. These tools were evaluated against a well-defined set of smart contracts for comparative analysis. Platforms like Oyente employ symbolic execution to simulate contract behaviour with symbolic inputs in order to traverse many execution paths, targeting specific vulnerabilities like reentrancy and the self destruct vulnerability. Tools like Mythril adopt hybrid approaches, employing concolic and taint analysis and control flow checking. On the other hand, Securify leverages static analysis to infer semantic information from contracts, ensuring derived patterns adhere to specific security properties. SmartCheck also employs static analysis, running lexical and syntactical analyses on Solidity source code. In section 4.1 of \cite{durieux_empirical_2020}, the challenges associated with these automated analysis tools are outlined. Despite the recent advances in these technologies, they delineated four pivotal areas requiring further research:

\begin{itemize}
    \item The quality of these tools requires an appreciable amelioration in performance, particularly concerning accuracy and false positives and negatives, which is paramount to the adoption and utility of such tools. 
    \item The breadth of vulnerabilities detected needs to be enlarged. There exists a great number of smart contract vulnerabilities. However, a compendium of definitions and testing sets encapsulating this compendium does not exist, resulting in variability and difficulties in comparing results in subsequent studies that employ different datasets.
    \item \cite{durieux_empirical_2020} argues that these analysis tools should be embedded in the development process, improving how developers interact with them, necessitating high accuracy levels and quick vulnerability analysis.
    \item There are limitations associated with the DASP10 taxonomy \cite{dasp10}, corroborating the need to include new categories reflecting recently discovered vulnerabilities. 
\end{itemize}

Given the predictive capabilities of machine learning to improve accuracy and learn features from data, our study implicitly imparts a focus on the first critique, thus necessitating a dedicated exploration into and synthesis of machine learning approaches in this area.

\subsection{Potential Dangers}

Despite many innovative advantages that smart contracts offer to automation and expediting transactions, the vulnerabilities mentioned above in these digitised agreements remain a perennial threat to the broader adoption and reliability of blockchain technologies. The consequences are not merely confined to financial losses but extend to broader systemic, pernicious risks such as diminished user trust \cite{zheng_overview_2020}. \cite{sayeed_smart_2020} delves into two famous hacks, highlighting the financial repercussions of improper security measures. The DAO attack in 2016 was a result of a recursive call bug, which was exploited to withdraw vast sums of money before the system could update the contract's balance, enabling the theft of millions of dollars worth of ether in hours. Similarly, the Parity Wallet hacks resulted from an exploitable bug permitting the use of the \textit{initWallet} function, enabling attackers to issue transactions claiming ownership of MultiSig wallets. These incidents stress the necessity of robust and efficient smart contract vulnerability detection mechanisms.



\section{Methodology} 
\label{sec:methodology}

Having detailed the background context and motivations behind our work, this section presents our research questions in details followed by  proposed framework. This flows into our search strategy, where we specify the databases we leverage to systematically search for relevant literature and explain the tailored search queries designed to retrieve the most relevant articles and papers. Following this, we establish well-defined exclusion criteria that will guide the screening and selection of studies, ensuring both relevance and methodological quality. To complement this process, a set of quality criteria is consulted to iteratively ensure that the papers are of a high academic calibre. Data extraction procedures are then outlined, specifying how essential information and findings will be distilled from the chosen articles for structured analyses. The selected papers are subsequently categorised according to which machine learning family they are a part of, ensuring a more organised, thematic literature review. Finally, we elucidate our data synthesis and analysis approach and propose potential limitations in our methodology. Our methodology was formulated by consulting seminal literature on systematic literature reviews \cite{kitchenham_2007, booth_systematic_2016, xiao_guidance_2019}.

\subsection{Research Questions}

To address current needs in the field, we have formulated  the following research questions to guide our investigation:

\begin{enumerate}

    \item[(RQ1)] \textit{What are the state-of-the-art machine learning techniques for identifying and mitigating specific vulnerabilities in smart contracts?} This question aims to assess the effectiveness of current machine learning techniques relative to traditional methods, laying the groundwork for the study.
    \item[(RQ2)] \textit{What machine learning algorithms have been applied to smart contract vulnerability detection, and how do they differ in efficacy and limitations when detecting vulnerabilities?} This question evaluates the different machine learning algorithms, interrelating the strengths and weaknesses of various research studies within each family of models.
    \item[(RQ3)] \textit{What are the current research gaps and future work opportunities in the application of machine learning for enhancing the detection and mitigation of vulnerabilities in smart contracts?} This question seeks to identify existing research gaps and suggests areas for future work.
    
\end{enumerate}

Collectively, these questions offer a comprehensive and well-rounded framework for exploring the field's academic and practical domains in order to produce an insightful scholarly resource.

\subsection{Search Strategy}

\subsubsection{Databases}

Per recommendations by \cite{kitchenham_2007}, we scrutinise five databases - namely IEEE Xplore, ACM Digital Library, ScienceDirect, Google Scholar and Scopus - for our systematic review. These databases were chosen to emphasise rigorously peer-reviewed academic journals and conference proceedings, facilitating a comprehensive exploration of pertinent empirical case analyses, methodological paradigms, and practical developments.

\subsubsection{Search Queries}

To capture the full scope of research, we devised a search query tailored to each database's search capabilities, targeting titles, abstracts, keywords, and full texts. The query employs the logical operators "AND" and "OR" and is explicitly phrased as: \\

\textbf{("smart contracts" OR "smart contract") AND ("machine learning" OR "deep learning" OR ("neural network" OR "neural networks")) } \\

This specificity restricts our focus to machine learning and deep learning, eliminating the ambiguity and generalisation linked with "artificial intelligence". Deep learning is included to explore specialised neural network architectures, while machine learning allows for a broader sweep of relevant algorithms. 

Additionally, an iterative adjustment was made to the query to address the limitations of its initial form. The singular versions of our key terms (i.e. smart contract, neural network) were expanded to include their plural counterparts. This modification resolved the issue of omitting relevant papers from the search results. This fine-tuning ensures that the literature retrieved is not only pertinent to the research objective but also inclusive of the full array of methodologies and insights in the targeted domain. We also decided to localise the search to focus primarily on abstracts. This targeted approach allowed for a more refined search output, as focusing on abstracts helped eliminate papers that were tangential or not directly aligned with the research objectives. By doing so, we ensured that the articles retrieved and subsequently analysed were highly germane, concentrated and relevant, fortifying the methodological rigour of our systematic literature review.

\subsubsection{Exclusion Criteria}

To maintain the focus and quality of this systematic literature review, certain types of articles and works will be excluded from consideration. The following outlines the exclusion criteria of our study:

\begin{itemize}
    \item Works that are not in English: Our review will be conducted in English and aims for a broad international audience. Therefore, articles published in other languages will be excluded. 
    \item Works that do not present any experimentation or empirical results: Articles that only provide theoretical propositions without empirical data, experimentation or comparative analysis will be excluded. For this review, it is critical that all included works are based on evidence-based findings. Further, if a work does not delve into the specifics of the machine learning model used, the paper is discarded.
    \item Duplicated articles across databases: If identical articles are found across databases, only one instance of the article will be included to avoid duplication and bias in the analysis.
    \item No institutional access: Papers that did not have institutional access or were not accessible through available channels were excluded from the review.
    \item Honeypot and Ponzi scheme-focused papers are not included, as these practices are more intentionally deceptive and malicious in nature than addressing technical vulnerabilities, which are seen more as unintended weaknesses of the code that can be exploited.  
\end{itemize}

Considering the contemporary nature of the research corpus, there will be no specific time frame requirements imposed for the inclusion of publications in the review.

\subsubsection{Quality Assessment}

While we acknowledge the importance of rigorously assessing the quality of individual studies, our review aims to present an all-encompassing panorama of the field. To ensure the inclusion of papers of solid quality, we undertook a streamlined quality assurance process focusing on the following criteria:

\begin{itemize}
    \item Peer review: Preference was given to articles published in peer-reviewed journals and conference proceedings.
    \item Reputable sources: papers from well-regarded journals or conferences were prioritised.
    \item Methodological transparency: Studies that transparently articulated their research design, methodology and data analysis were given higher consideration.
    \item Alignment with research questions: Only papers that directly contributed to answering our research questions were included, implicitly satisfying a criterion of relevance and focus. Those that were only tangentially related were discarded.
\end{itemize}

Although this approach may deviate from more traditional, in-depth quality assessments, it suits the comprehensive nature of our research questions. Our review aspires to extract valuable insights from a broad range of studies, acknowledging that even studies of comparatively lower methodological rigour may offer unique, exploratory 'nuggets' of information that contribute to a more complete understanding of our research area \cite{booth_systematic_2016}, permitting an exhaustive approach to the review and a balanced aggregation of the existing body of knowledge. It is through this lens that our quality assurance process should be interpreted.

\subsubsection{Data Extraction}

The data extraction process is structured to maintain the reliability and comprehensibility of information collected from each selected paper. All extracted data will be populated into a comprehensive matrix to ensure uniformity in the extraction process. The matrix will contain columns corresponding to the following pivotal fields:

\begin{itemize}
    \item Main contributions: Encapsulates what the paper adds to the existing corpus of knowledge, whether it introduces a new method or originally fuses existing techniques. 
    \item Machine learning techniques and methodologies: Concentrates on documenting the machine learning methods articulated in the paper.
    \item Vulnerabilities targeted: Identify the types of smart contract vulnerabilities the paper seeks to detect or mitigate. A standardised lexicon of vulnerabilities will be developed to ensure coherence across the collected data.
    \item Limitations, critiques and future work: Notes limitations, pitfalls of the paper and highlights areas requiring future inquiry.
\end{itemize}

Each row in the matrix will correspond to a specific paper, ensuring a structured and organised repository of the extracted information. The "contributions" and "limitations, critiques and future work" components of the matrix will be synthesised in our results and discussion, whereas the "machine learning techniques and methodologies" and "vulnerabilities targeted" components will be published as separate tables in the appendix of this paper. 

\subsubsection{Paper Categorisation}

In this stage, the selected papers will be sorted based on the type of machine learning technique employed. This approach to categorisation allows for an understanding of how specific models contribute to the goal of detecting and mitigating vulnerabilities in smart contracts. It facilitates a granular comparison of efficacy across diverse models. This enriches the review's qualitative depth by revealing performance distinctions that may be model-specific. By organising the papers based on model type, we also aim to identify prevailing trends, opportunities, and limitations in each category. Furthermore, this categorisation can serve as an analytical tool for future research, providing a clear understanding of the types of machine learning techniques that are promising and under-explored in the context of smart contracts. It can help researchers and practitioners more efficiently navigate the complex landscape of existing literature, accelerating advancements in the field.

\subsubsection{Synthesis of Results}

In this part of the process, we aim to aggregate and summarise the machine learning techniques employed, the effectiveness of these techniques, and the vulnerabilities addressed in the collected pool of papers to derive insights about the current experimental landscape. The synthesised data will be displayed through various graphs, tables and charts. This data will be analysed, and the papers comprising the repository of works will be scrutinised in depth.

\subsubsection{Limitations}

While our methodology aims to be comprehensive, it is important to acknowledge certain limitations that could potentially impact the interpretation and universality of our findings:

\begin{itemize}
    \item Publication bias: Our insights may skew more towards authoritative works in the field, limiting representation.
    \item Inconsistent metrics: Variability in effectiveness metrics across studies may complicate direct comparisons.
    \item Study quality: Included studies may have biases or methodological flaws, impacting reliability.
    \item Method complexity: Some machine learning methods' complexity adds intricacies to consider during interpretation, as well as classification in our "Paper Classification" process.
    \item Inconsistencies in vulnerability definitions and datasets: Lack of a standardised classification of vulnerabilities could complicate data synthesis.
    \item General search query: While our search ensures an exhaustive coverage of the literature, it may involve a substantial body to siphon through in order to find all pertinent papers.
\end{itemize}

\section{Results} \label{sec:results}

In this section, we outline the results of our systematic literature review. We delineate the process of collecting and refining our sources and illustrate the distribution of publications over time. We also categorise our machine learning models to identify the common approaches to vulnerability detection and explore the popularity of the vulnerabilities being researched. In so doing, we address our first research question \textbf{(RQ1)}: What are the state-of-the-art machine learning techniques for identifying and mitigating specific vulnerabilities in smart contracts? Finally, we hone in on the assortment of machine learning techniques in synthesising our results, diving into how each of these papers contributes to the research of machine learning applications to smart contract vulnerability detection.

\subsection{Search Results}

\begin{figure}[!htbp]
\centering
\includegraphics[width=\textwidth]{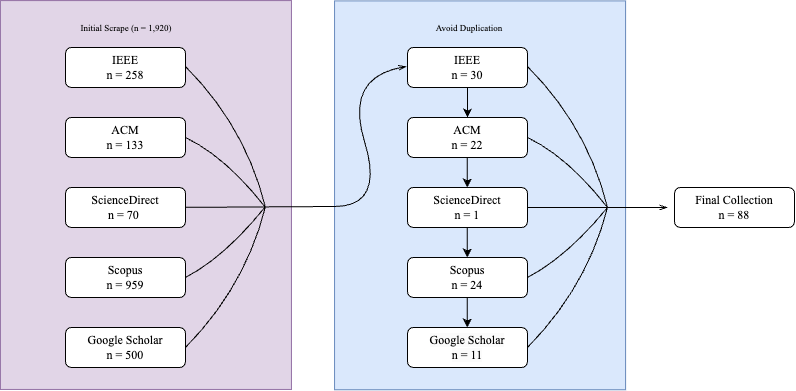}
\caption{Flowchart of the search strategy.}
\label{fig:search_strategy_flowchart}
\end{figure}

The search operations were executed on September 10, 2023. Our search strategy, depicted in figure \ref{fig:search_strategy_flowchart}, yielded an initial pool of 1,920 sources. This pool was comprised as follows: IEEE $(n = 258)$; ACM $(n = 133)$; ScienceDirect $(n = 70)$; Scopus $(n = 959)$; Google Scholar $(n = 500)$. Following the inclusion and exclusion criteria tailored to the research questions, our final selection of studies was narrowed down to $n = 88$ sources by scrutinising the collections pulled from each database in a linear fashion to avoid adding duplicates to the final pool. This final collection was comprised as follows: IEEE $(n = 30)$; ACM $(n = 22)$; ScienceDirect $(n = 1)$; Scopus $(n = 24)$; Google Scholar $(n = 11)$. The considerable reduction in papers is indicative of the specificity of our research criteria and the novelty of the research domain. 

\subsection{Year of Publication Distribution}

\begin{figure}[htbp]
\centering
\includegraphics[width=\textwidth]{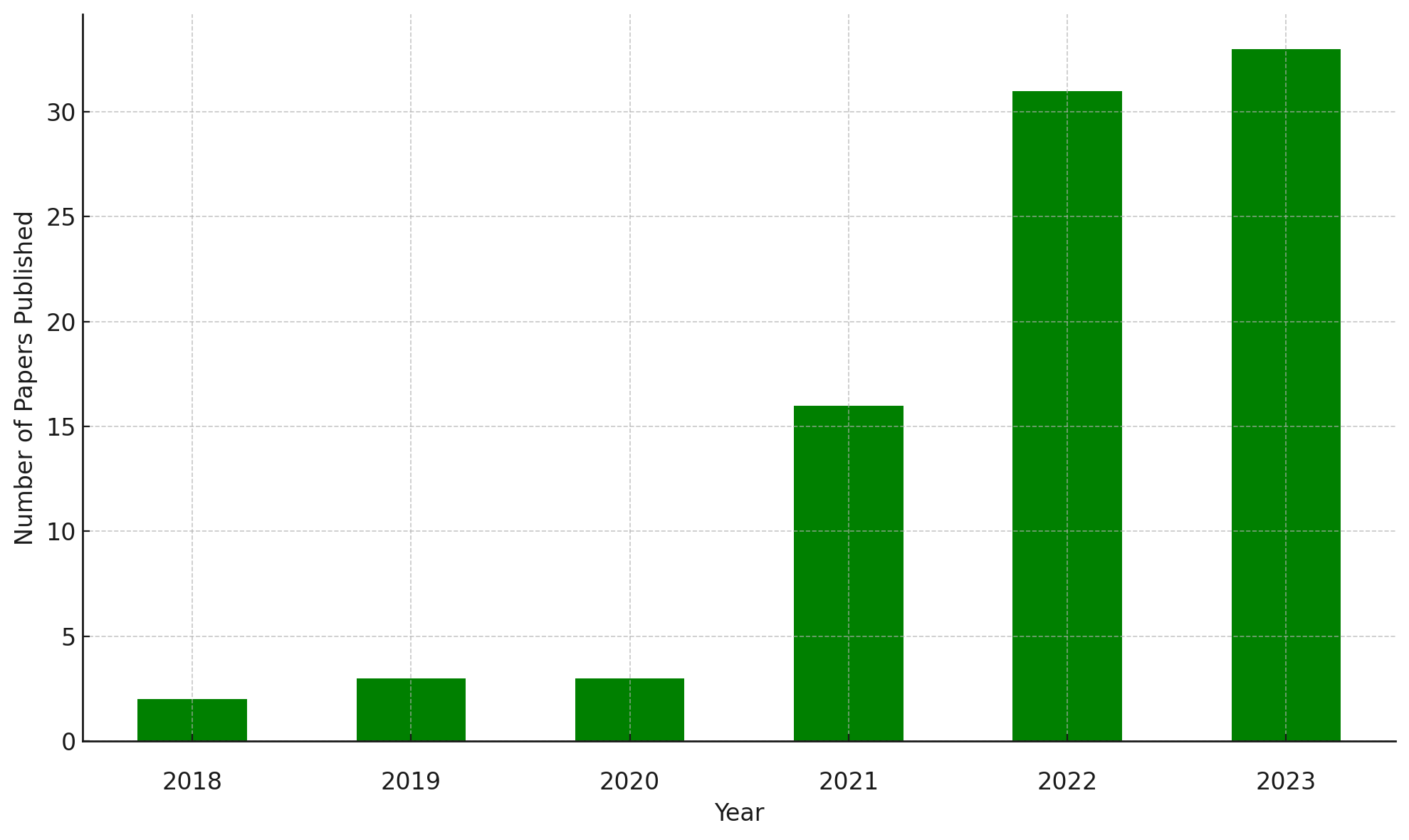}
\caption{Number of papers published each year.}
\label{fig:papers_per_year}
\end{figure}

Analysing figure \ref{fig:papers_per_year}, given the databases we have selected, 2018 is the first year in which research into the employment of machine learning to address smart contract vulnerabilities begins, with only two publications that year. There was a significant jump to 31 published papers in 2022 and 34 published papers so far in 2023, suggesting a strong focus in this research area has germinated. Given the increasing trend, more recent works may reflect the current state of research. In order to better understand the evolution of this field, we pay particular attention to these years to synthesise the most current techniques and practical models.

\subsection{Range of Machine Learning Models}

{
\setstretch{1}
\renewcommand{\arraystretch}{1.5}
\small 
\begin{longtable}{p{3cm} p{9cm}}
    \caption{Machine Learning Model Nomenclature}
    \label{fig:nomenclature}
    \\
    \toprule
    Abbreviation & Full Name \\
    \midrule
    \endfirsthead
    \multicolumn{2}{c}{\tablename\ \thetable{} -- Continued} \\
    \toprule
    Abbreviation & Full Name \\
    \midrule
    \endhead
    \bottomrule
    \multicolumn{2}{r}{} \\
    \endfoot
    \bottomrule
    \endlastfoot
    \begin{filecontents*}{csvs/nomenclature.csv}
Abbreviation,Full Name
AdaBoost,Adaptive Boosting
ANN,Artifical Neural Network
AWD-LSTM,ASGD Weight-Dropped LSTM
BAL,Bayesian Active Learning
BERT,Bidirectional Encoder Representations from Transformers
BiGRU,Bidirectional GRU
BiLSTM,Bidirectional LSTM
BRKNN,Binary Relevance KNN
CNN,Convolutional Neural Network
DT,Decision Tree
DCN,Deep Cross Network
DL,Deep Learning
DNN,Deep Neural Network
DR-GCN,Degree-Free GCN
DA-GCN,Dual-Attention GCN
EEC,EasyEnsembleClassifier
XGBoost,Extreme Gradient Boosting
GRU,Gated Recurrent Unit
GB,Gradient Boosting
GAT ,Graph Attention Network
GCN,Graph Convolutional Network
GMN,Graph Matching Network
GNN,Graph Neural Network
HGT,Heterogeneous Graph Transformer
HAM,Hierarchical Attention Mechanism
KNN,K-Nearest Neighbours
L,Linear
LR,Logistic Regression
LSTM,Long Short-Term Memory
MRN-GCN,Multi-Relational Nested Graph Convolutional Network
MLKNN,Multilabel KNN
MODNN,Multiple-Objective Detection Neural Network
NB,Naïve Bayes
NN,Neural Network
P,Perceptron
RDF,Random Deep Forest
RF,Random Forest
RNN,Recurrent Neural Network
RGCN,Residual Graph Convolutional Neural Network
SBCNN,Serial-Parallel Convolutional Neural Network (SPCNN) 
SGD,Stochastic Gradient Descent
SVM,Support Vector Machine
TCN,Temporal Convolutional Network
TMP,Temporal Message Propagation
TextCNN,Text Convolutional Neural Network
TBCNN,Tree-Based Convolutional Neural Network
xATT,x with Attention Mechanism
\end{filecontents*}
\csvreader[
        late after line=\\,
        before reading={\catcode`\_=12},
        after reading={\catcode`\_=8},
        respect underscore
    ]
    {csvs/nomenclature.csv}{}
    {\csvcoli & \csvcolii}
\end{longtable}
}

\begin{figure}[htbp]
\centering
\includegraphics[width=0.7\textwidth]{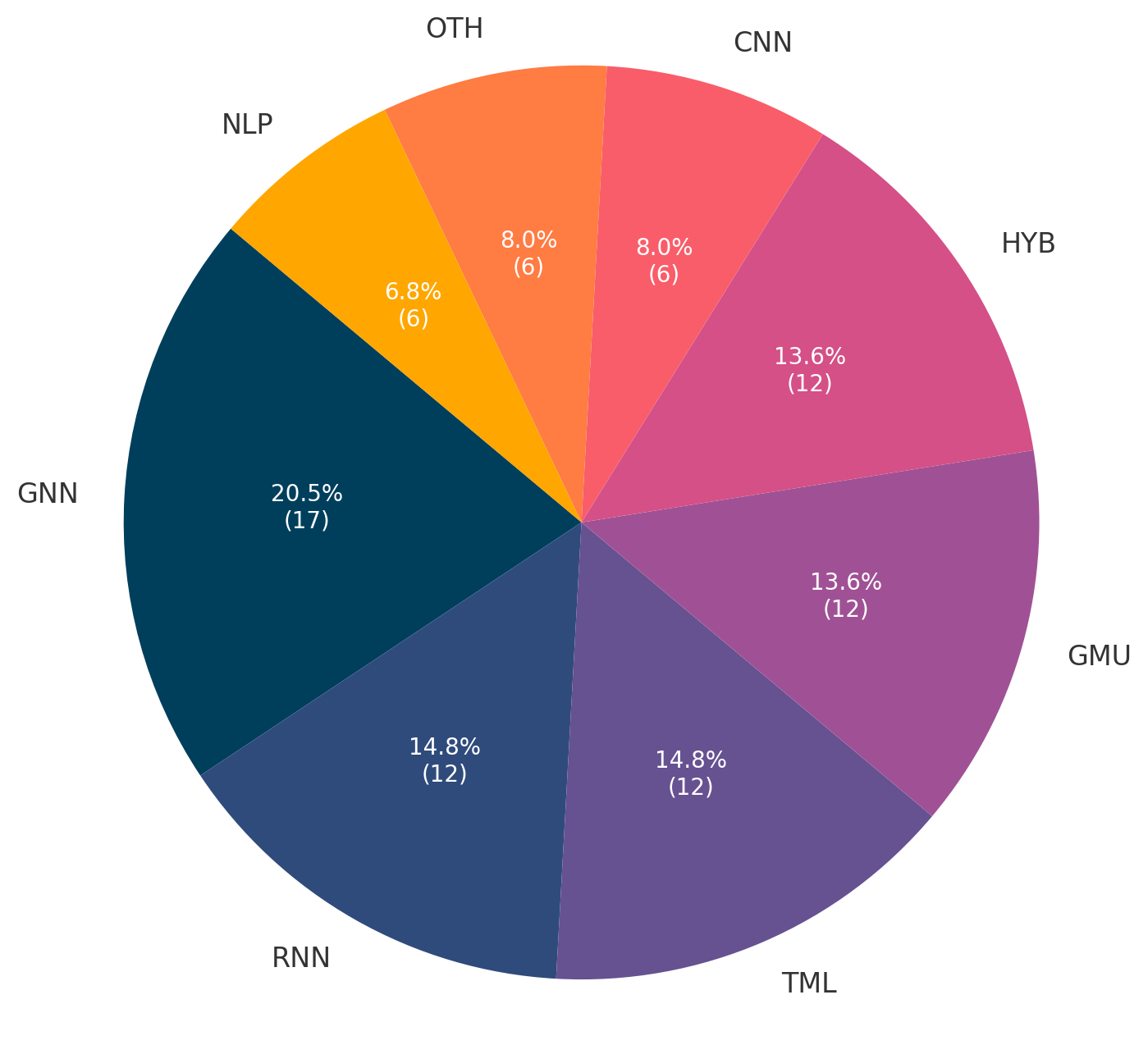}
\caption{Presence of machine learning models in the literature.}
\label{fig:category_distribution}
\end{figure}

\begin{figure}[htbp]
\centering
\includegraphics[width=0.6\textwidth]{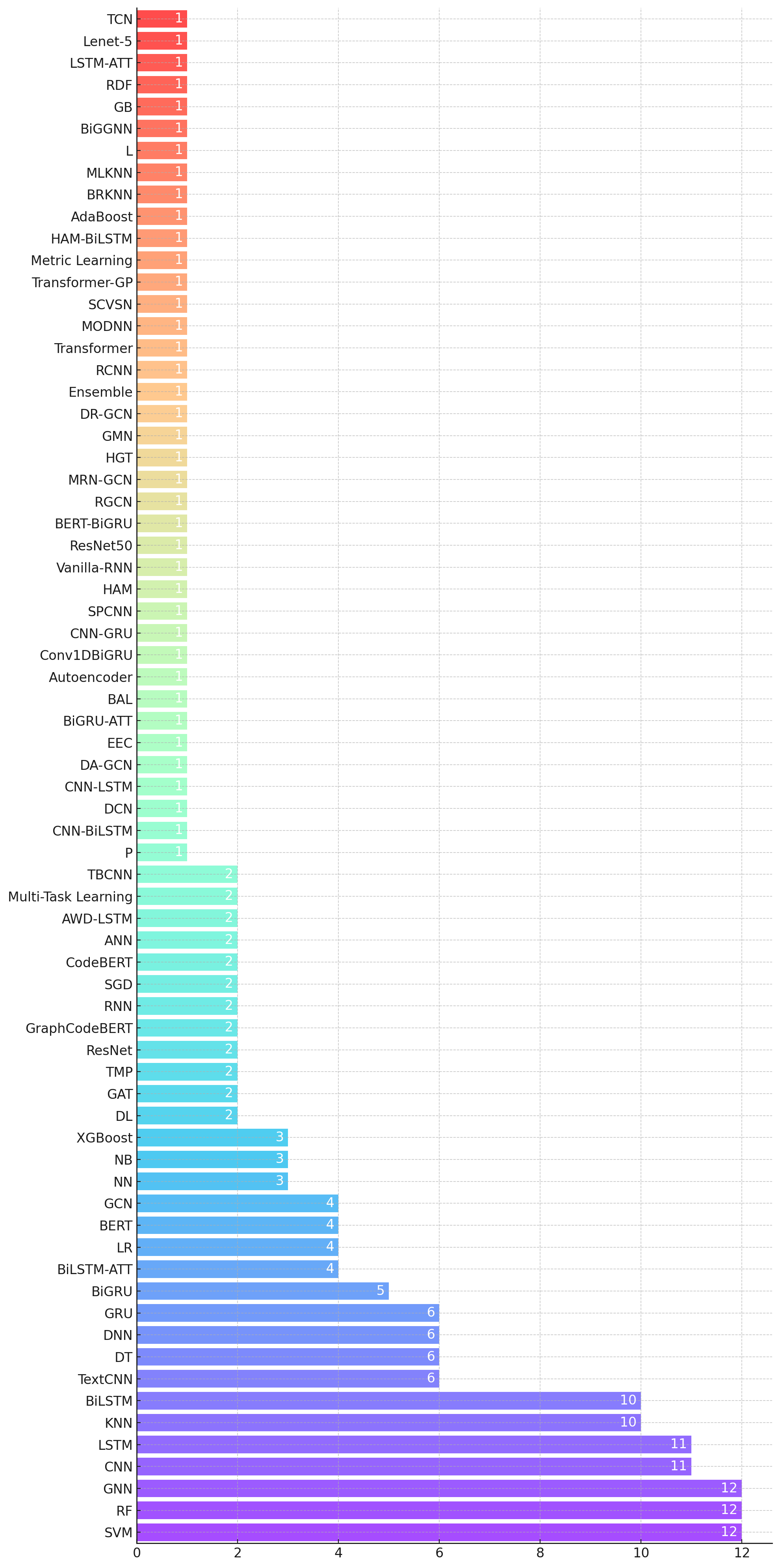}
\caption{Presence of machine learning models in the literature.}
\label{fig:model_occurrence}
\end{figure}

In table \ref{fig:nomenclature}, we have compiled an index of nomenclature to complement our later discussions of the literature. In consulting this table, as well as figure \ref{fig:model_occurrence}, we see the great diversity in machine learning approaches being employed in enhancing smart contract security. GNN, SVM, and RF were the most frequently occurring models in our dataset, each being primarily employed in some sense 12 times. CNN and LSTM follow closely, with 11 uses each. The dataset includes 70 different models, the mix of models suggestive of the existence of a myriad of machine learning approaches and techniques in the enhancement of smart contract security. In particular, the popularity of CNNs, LSTMs and GNNs suggests the necessity of models capable of handling complex feature extraction capabilities. In contrast, more traditional machine learning models like SVM and RF may be employed due to their universality of application across a range of classification problems. There are several models mentioned only once. In these cases, it appears research is being undertaken to experiment with some specialised variant or hybridisation of previous models to probe their efficacy in addressing the problem at hand, and the models are not yet widely adopted in this specific domain. On the surface, this variety indicates a level of fragmentation in the models being researched. In order to observe broader trends and draw more meaningful comparisons between papers, we classify each paper according to a set of model groups based on which model they primarily use. The classifications are detailed in \ref{fig:classification}, and the classifications of all papers can be found in the repository in figure \ref{fig:papers_repository}. The frequencies of each classification are also depicted in figure \ref{fig:category_distribution}. Referring to the classification distribution, we see that the research landscape is somewhat fairly distributed.

\vspace{0.5cm}

{
 \setstretch{1}
\renewcommand{\arraystretch}{1.5}
\small 
\begin{longtable}{p{1.5cm} p{6cm}}
    \caption{Broader Paper Classification} 
    \label{fig:classification} \\
    \toprule
    Abb. & Title \\
    \midrule
    \endfirsthead
    \multicolumn{2}{c}{\tablename\ \thetable{} -- Continued} \\
    \toprule
    Abb. & Title \\
    \midrule
    \endhead
    \bottomrule
    \endfoot
    \bottomrule
    \endlastfoot
    CNN & Convolutional Neural Networks \\
    GMU & General Multi-Model Usage \\
    GNN & Graph Neural Networks \\
    HYB & Hybrid Models \\
    NLP & Natural Language Processing \\
    OTH & Other Models and Techniques \\
    RNN & Recurrent Neural Networks \\
    TML & Traditional Machine Learning Techniques \\
\end{longtable}
}

\vspace{2cm}

\subsection{Vulnerability Addressal}

\begin{figure}[htbp]
\centering
\includegraphics[width=\textwidth]{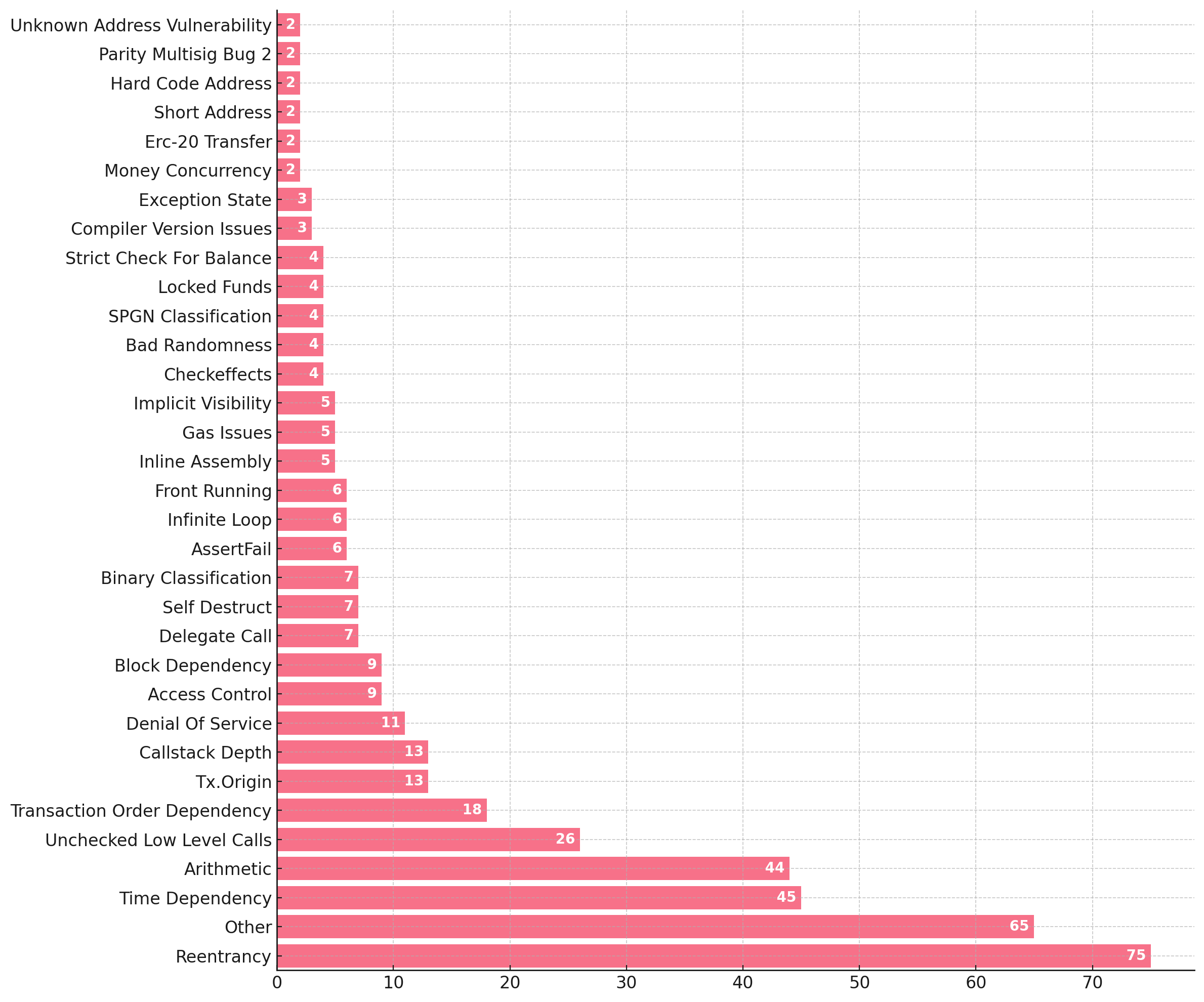}
\caption{Count of vulnerabilities addressed over our collection of papers.}
\label{fig:vulnerability_count}
\end{figure}

Figure \ref{fig:vulnerability_count} paints a picture of which vulnerabilities are most popularly researched.  Reentrancy vulnerabilities are by far and away the most investigated vulnerability, with 75 papers addressing them. Timestamp dependency and arithmetic vulnerabilities (such as integer overflow and underflow) follow with 45 and 44 occurrences, respectively. Unchecked low level calls and transaction order dependency vulnerabilities were also significant, with 26 and 18 occurrences, respectively. These vulnerabilities are among the most critical to address, substantiating their prevalence as a focus for many of these papers. The "Other" category is comprised of vulnerabilities that were targeted only once. Whilst some vulnerabilities are targeted relatively incessantly, the precipitous drop in focus on other vulnerabilities is noticeable, especially given the large number of "Other" vulnerabilities. Four papers adopted a classification (SPGN) derived from \cite{nikolic_finding_2018}'s work, which partitions the vulnerabilities into four severity levels: greedy, prodigal, suicidal and normal. Some papers also simply perform a binary classification as to whether the smart contract contains a vulnerability and do not explicitly suggest what type of vulnerability it is. Evidently, there is not only a large number of vulnerabilities addressed in our collection of research but also several ways in which classification and definition of vulnerabilities can be performed.


\section{Synthesis of Results} \label{sec:synResults}

In this section, we synthesise our results and dive deeper into the papers returned from our search. In so doing, we address our second research question \textbf{(RQ2)}: What machine learning algorithms have been applied to smart contract vulnerability detection, and how do they differ in efficacy and limitations when detecting vulnerabilities? By delving into the effectiveness of each proposed method, we observe that whilst classical machine learning algorithms offer advantages in speed, more complex and contemporary models offer advantages in accuracy, a pitfall of traditional analysis tools mentioned in section \ref{sec:existing_tools}.

\subsection{Traditional Machine Learning Techniques}

\cite{xu_novel_2021} introduces a vulnerability detection model based on Abstract-Syntax-Tree analysis. The framework automatically extracts structural similarities from ASTs to build feature vectors, eliminating the necessity for manual intervention. KNN is employed to predict eight types of vulnerabilities, producing 90\% along the accuracy, recall and precision metrics. These results outperformed traditional vulnerability detection methods like Oyente and SmartCheck in accuracy and training time. Similarly, \cite{yang_smart_2022} introduce an approach using ASTs for automatic feature extraction. Their approach divides ASTs into sub-ASTS based on function types, state variables and function modifiers. They then utilise the transformed feature set to train RF models, reporting a high model performance with an average accuracy of 98.7\% and an F1-value of 88.5\% in detecting three types of vulnerabilities. 

\cite{mandloi_machine_2022}, \cite{eshghie_dynamic_2021} and \cite{momeni_machine_2019} put forward frameworks which employ RF and DT classifiers. \cite{mandloi_machine_2022} proposes a framework that dynamically monitors threats using transaction meta-data and balance data from the blockchain. They utilise machine learning classifiers, namely RF and DT, to identify transactions as either agreeable or unfavourable and achieve a cumulative performance accuracy of 98\% in tests involving 540 transactions. \cite{eshghie_dynamic_2021} proposes Dynamit, a monitoring framework for detecting reentrancy vulnerabilities. The novelty of their paper lies in the reliance on transaction metadata and balance data, eliminating the need for domain knowledge, code instrumentation, or a special execution environment. The framework extracts features from transaction data and employs a machine learning model to classify transactions as benign or harmful. Not only does it identify vulnerable contracts, but it provides an execution trace of the attack. The framework has an accuracy of over 90\% on 105 transactions using RF.
\cite{momeni_machine_2019} adapts two static code analysers, Slither and Mythril, to label over 1,000 verified Ethereum smart contracts and extract 17 features from the dataset indicative of vulnerabilities. They utilised four supervised binary classifiers (SVM, RF, DT, NN). They achieved an average accuracy of 95\% and an F1 score of 79\%, arguing for an approach that harnesses the strengths of static code analysers but delivers results more efficiently. \cite{hao_scscan_2020} introduces a comprehensive SVM-based scanning system covering six vulnerabilities, which achieves an identification of over 90\% for several popular attacks. \cite{yang_improvement_2023} also compares their work, SCSVM, an SVM model based on meta-learning, to other static analysis tools like Mythril, Oyente and Slither, demonstrating superior performance to these tools. 

\cite{wang_contractward_2021} and \cite{lakshminarayana_towards_2022} employ more advanced classical machine learning classifiers, such as XGBoost. \cite{wang_contractward_2021} proposed ContractWard, offered as an alternative to time-consuming symbolic execution methods. They employ a similar methodology to \cite{song_efficient_2019}, collecting 49,502 smart contracts for training and testing and creating a 1,619-dimensional feature space using extracted bigram features. They used algorithms such as XGBoost, AdaBoost, RF, SVM and KNN for comparative analysis and achieved Micro-F1 and Macro-F1 scores of over 96\% using XGBoost, with its detection time significantly outperforming Oyente and Security; 4 seconds as opposed to 28 seconds and 18 seconds respectively. \cite{lakshminarayana_towards_2022} utilises the XGBoost machine learning model to achieve an average 2\% improvement in F1 score results for detecting four smart contract vulnerabilities. They employ Business Process Modeling notation to create graphical representations of smart contracts for better stakeholder understanding, addressing a significant limitation of smart contracts: lack of comprehensibility for non-technical stakeholders.

\cite{zhao_novel_2023} posits a multi-faceted feature extraction strategy that fuses information from abstract syntax trees, opcodes, and control flow graph basic blocks, improving the completeness and accuracy of the features used for vulnerability detection. They also developed an automated approach for dataset labelling employing code embedding and similarity comparison technologies to address the inefficiencies related to manual labelling. Using the Smartbugs and SolidiFi-benchmark datasets alongside six machine learning algorithms (KNN, SVM, DT, RDF, K, P), they conducted extensive experiments, demonstrating superior performance over existing models in detecting reentrancy vulnerabilities. \cite{song_efficient_2019}’s model, which extracted 1,619-dimensional bigram features from simplified operation codes to build a comprehensive feature space, achieved over 96\% in predictive recall and precision. The model also bore a good time efficiency, with an average detection time of about 4 seconds per contract, making it suitable for batch processing. The model is grounded in the RF algorithm and is compared to SVM and KNN for benchmarking. \cite{li_detecting_2023} also invented a machine learning model combining the n-gram model, a generalised form of the bigram model, and a vector weight penalty for feature extraction from opcode sequences. They validated this model using the KNN, SVM, and LR models, with SVM achieving a 91.4\% accuracy and 75.3\% F1-score. \cite{lohith_tp-detect_2023} introduces a classification method using several machine learning models: MLkNN, BRkNN, RF, and NB. Their dataset was constructed via pixel values from images and trigram feature extraction. They find that NB outperforms all other algorithms with F1-scores of 99.38\% and 99.44\% in their binary relevance and classifier chain setups, demonstrating the superiority of the TriPix dataset over other approaches like opcode characteristics or image-based detection.

\subsection{Neural Networks and Deep Learning}
Neural networks are designed similarly to how our brains are composed. Neural networks detect patterns and features through feature learning and can model complex and non-linear relationships, making them more suitable for tasks with large and complex datasets. 

Deep Learning is a sub branch of machine learning that utilises multi-layered neural networks, with at least three or more layer. This deeper architecture enables capturing patterns between within the data.

\subsubsection{General Multi-Model Usage}

Several papers in this category propose feature extraction and optimisation techniques. \cite{xing_new_2020} introduces a novel feature extraction method called "slicing matrix”. They conducted experiments to compare the efficacy of neural networks combined with slicing matrix versus neural networks combined with opcode features for vulnerability detection. They demonstrated that the use of slicing matrix enhances the accuracy of vulnerability detection in smart contracts. They identified that the Random Forest-based model using opcode features outperformed other models in detecting smart contract vulnerabilities. \cite{yu_deescvhunter_2021} proposes DeeSCVHunter, a modular and systematic deep-learning framework. The paper introduces Vulnerability Candidate Slice (VCS), a novel feature that encapsulates richer syntax and semantic information for improving vulnerability detection. They achieve up to 25.76\% improvement in F1 score through the incorporation of VCS and outperform seven state-of-the-art vulnerability detection methods in experiments conducted on real-world datasets. \cite{wu_detecting_2022} proposed the use of opcodes as static features and applied bigram and Mtfidf optimization, utilising CNN, LSTM, CNN-BiLSTM, and ResNets for vulnerability detection. They achieved promising results, with ResNets reaching an 82\% Macro-F1 score, and highlight the superiority of deep learning methods over traditional machine learning for vulnerability detection.

The majority of papers in this category use a selection of deep learning models that couldn’t be directly assigned to one of the other categories we have defined. \cite{lutz_escort_2021} posit ESCORT, a DNN-based framework that supports transfer learning and multi-vulnerability detection. The framework operates on bytecode and distinguishes safe from vulnerable contracts, identifying vulnerability types. ESCORT achieved an average F1 score of 95\% and is generalisable. \cite{ashizawa_eth2vec_2021} proposes Eth2Vec, a tool for detecting vulnerabilities that use neural networks driven by the Paragraph Vector Distributed Memory (PV-DM) model to automatically learn relevant features from EVM bytecode. The model can quickly analyze contracts, taking an average of 1.2 seconds per contract, and achieves 86.6\% precision in detecting critical reentrancy vulnerabilities, surpassing existing Word2Vec tools. \cite{li_smart_2022} proposes Link-DC, a modular vulnerability detection model that utilises contract graphs and expert-defined pattern features as input data rather than treating smart contract source code as simple text sequences. The paper employs deep and cross-networks to transform low-dimensional, sparse features into high-dimensional, nonlinear features. The work achieves high detection accuracy for specific vulnerabilities: reentrancy (94.37\%), timestamp-dependent (92.11\%), and infinite loop (85.29\%). \cite{deng_smart_2023} combines deep learning and multimodal decision fusion, achieving high accuracy and recall rates for multiple vulnerability types. The paper utilises multiple features extracted from the source code, operation code, and control-flow graph of smart contracts for comprehensive vulnerability detection without reliance on expert-defined rules. The paper also offers a comprehensive examination of the impact of different modalities (source code, operation code, control-flow graph) on the task of vulnerability detection, contributing to the understanding of the strengths and weaknesses of each.

\cite{yuan_svchecker_2022} proposes a method for extracting specific code snippets from Solidity source code, labelling them as malicious or benign, called SVChecker. They employ a neural network with a transformer-encoder and a multi-head self-attention mechanism. The paper compares the efficacy of SVChecker against traditional methods and existing tools, with results indicating that SVChecker achieves higher precision in vulnerability detection. \cite{mi_vscl_2021} introduces VSCL, a framework for automatic detection that leverages bytecode analysis and a metric learning-based deep neural network. They validate their approach using a large-scale Ethereum smart contract benchmark, demonstrating superior performance over existing methods. \cite{gopali_vulnerability_2022}’s model categorises known vulnerabilities into multi-class categories, utilising LSTM and TCN to detect and classify vulnerabilities. They find that the TCN-based model outperforms existing work, including LSTM models, and reports high-performance metrics for the TCN model: 95.95\% precision, 93.73\% recall, 94.83\% F1 score, and 94.77\% accuracy.

There were also a number of fuzz testing approaches. \cite{liao_soliaudit_2019} posits SoliAudit, which combines machine learning and fuzz testing for assessing vulnerabilities in smart contracts. Their model is able to identify 13 types of vulnerabilities in Solidity code, achieving up to 90\% accuracy on 17,979 samples. They also implement a grey-box fuzz testing mechanism consisting of a fuzzer contract and a simulated blockchain environment for online transaction verification. \cite{xue_xfuzz_2022}’s machine learning-based approach xFuzz aims at reducing the search space of exploitable paths, achieving a 95\% recall on a testing dataset of 100K contracts.

subsection{Convolutional Neural Networks}

\cite{zhou_vulnerability_2022} proposes a Tree-based Machine Learning Vulnerability Detection (TMLVD) method, which uses abstract syntax trees (AST) for vulnerability analysis. They utilise a specialised convolutional network to extract features, enhancing vulnerability detection in smart contracts. They also conduct experiments that indicate TMLVS outperforms existing benchmarks in speed and accuracy. \cite{xu_efficient_2023} introduces a deep learning-based static method by reformulating the vulnerability detection problem into a natural language text classification task. Adapting a SkipGram model, they map smart contract codes to a unified vector space and train using a TextCNN model. They implement a code obfuscation technique to enhance the dataset and make the model more robust to redundant code interference. They also propose three opcode sequence generation methods to construct features and find that larger sample sizes lead to better feature extraction and higher detection accuracy. Their method outperformed SoliAudit with an average prediction accuracy of over 96\% and a detection time of less than 0.1 seconds. \cite{sun_attention-based_2021} proposes a CNN combined with a self-attention mechanism to reduce the need for manual analysis. Their method incorporates an enhanced one-hot encoder, sensitive word sharding, and a stop word list into their feature engineering process to improve the model’s performance. They develop a general benchmark for vulnerability detection techniques to reduce the requirement for specialised expertise. They suggest that their model can serve as a rapid pre-filter for traditional symbolic analysis methods to increase the overall detection accuracy and efficiency. \cite{huang_hunting_2018}’s model also focuses on reducing the need for manual analysis by using a CNN for automated feature extraction and learning. They translate the bytecode of Solidity smart contracts into RGB colour codes and convert these codes into fixed-sized encoded images to detect compiler bugs post-training. CodeNet, a CNN architecture model designed by \cite{hwang_codenet_2022}, aims to maintain the semantics and contract of smart contracts during vulnerability detection. They posit a unique data pre-processing technique retaining the locality of smart contract code to ensure that the transformed input images for CNN analysis preserve essential code features. \cite{hwang_codenet_2022} demonstrates through experimental evaluations that CodeNet’s detection performance outperforms six existing state-of-the-art vulnerability detection tools in performance and detection time. \cite{rossini_smart_2022} investigates various neural networks for classifying smart contract vulnerabilities. They compare four neural network types: LSTM, ResNet 1D CNN, 2D ResNet-18 CNN, and 2D Inception v3 CNN. They found that ResNet 1D CNN on bytecode showed the best classification results and that CNNs demonstrate good classification accuracy. 
\cite{qin_smart_2022}’s work employs a Critical Combination Path (CCP) methodology to isolate vulnerability-related code, thereby reducing extraneous noise in smart contracts. After analysing contract characteristics, they propose a code normalisation technique that minimises noise by eliminating homogeneous assembly code. The model utilises the normalised CCP, vectorises it with SimHash, and converts it to greyscale images for neural network classification. It is validated through experiments on real Ethereum smart contract datasets, showing improved performance over existing deep learning models.

\subsubsection{Recurrent Neural Networks}

\cite{gupta_deep_2022} introduces a deep learning-based methodology for detecting vulnerabilities in smart contracts, utilising ANN, LSTM, and GRU models. They conducted a comparative analysis between the three models - LSTM was found to outperform the others. \cite{gupta_deep_2022} also implements a simulation to classify smart contracts using the LSTM model prior to their blockchain deployment, propose a reward and penalty scheme based on smart contract classification, and provide evidence for the system’s data storage cost-efficiency and scalability, substantiating its utility in a blockchain environment. \cite{wang_ethereum_2023} introduced a unique framework that uses metric learning for optimising feature representation. They enhance the interpretability of their model by identifying critical vulnerability features through word vectorisation and attention mechanisms. Their BiLSTM model with attention mechanism outperformed the standard BiLSTM model and maintains an approximate 90\% accuracy in several vulnerabilities. The model also outperforms other deep learning models like TextCNN, RNN, LSTM-ATT, GRU, CodeBERT, and XLNet in terms of accuracy and precision. \cite{gogineni_multi-class_2020} proposes the use of an AWD-LSTM for classifying smart contracts into four categories: Suicidal, Prodigal, Greedy or Normal. Their architecture combines a pre-trained encode with a ‘custom head’ for better multi-class classification. The model achieves an accuracy of 91.0\% and an F1 score of 90.0\% in multi-class classification, indicating robust performance. \cite{qian_multi-label_2022} establishes a multi-label classification model for smart contract opcodes, improving upon existing models which only perform binary classification of a single vulnerability type. Their neural network architecture consisted of a BiLSTM layer, an attention layer and a fully connected layer. It achieved a performance exceeding 85\% across four evaluation metrics and for the detection of each of the five different smart contract vulnerabilities. \cite{xu_reentrancy_2022} floats a HAM-BiLSTM, which achieved higher detection accuracy and lower false alarm rates compared to existing methods. They found that account information of smart contracts also contains important information that can have an impact on the identification of vulnerabilities in smart contracts. Blass, an approach using BiLSTM-ATT for vulnerability detection, is put forth in \cite{ren_smart_2023}. They devise an algorithm for Complete Program Slices with semantic structure. They also introduced a code chain method for enhanced vulnerability capture and integrated an attention mechanism in the neural network to improve performance. \cite{zhang_smart_2022} introduces a BiLSTM model for smart contract vulnerability detection, which achieves high defect detection rates in practical scenarios, reducing human error and manual analysis. However, their model’s performance in terms of recall is suboptimal. \cite{zhou_smart_2023} employs a BiLSTM model in their framework, introducing mixed parameter sharing in a multi-task learning environment and adding unique fusion modes for task-specific sharing. Employing this technique, their model improves vulnerability detection precision, recall and F1 score compared to a standalone BiLSTM model. \cite{hu_smart-contract_2023} proposes a combined use of opcode sequences and ASTs for parsing smart contract code and employ a neural network with BiLSTM and a self-attention mechanism. The complementary parsing methods proved beneficial in boosting performance, particularly in detecting reentrancy vulnerabilities. \cite{goswami_tokencheck_2021} proposes an LSTM neural network model for vulnerability detection in ERC-20 contracts, achieving a 93.26\% accuracy and outperforming existing symbolic analysis tools in terms of time efficiency, confirming its effectiveness and positioning it as a viable alternative to existing methods. \cite{qian_towards_2020} uses a BiLSTM-ATT for precise reentrancy bug detection and introduces “contract snippet representations” for better semantic capture, outperforming state-of-the-art methods like Oyente, Securify, Smartcheck and Mythril. \cite{tann_towards_2018} uses an LSTM model that surpasses the Maian symbolic analysis tool, also noting that the LSTM model maintains a constant analysis time even as smart contract complexity increases, ensuring scalability.  The paper also validates that LSTM’s performance escalates with the introduction of new contracts. \cite{qian_bilstm-attention_2022} proposes a highly time-efficient model using a BiLSTM-Attention, which treats opcodes as sentences. Their paper aims to find the optimal opcode length to optimise detection accuracy, which turns out to be 6000, achieving a detection accuracy of 95.4\%, outperforming single and combined models in performance.

\subsubsection{Graph Neural Networks}

\cite{wang_gvd-net_2022} introduces GVD-net, a three-section model for smart contract vulnerability detection utilising Control Flow Graphs and Node2Vec for feature extraction. They establish a threshold for classifying contracts as safe or dangerous and achieve an accuracy of 90.2\% on the SBcurated dataset.  \cite{liu_smart_2023} introduces a two-phase smart contract vulnerability detection mechanism that integrates a GNN model and expert rules to enhance detection, efficiency, and generalisation. They developed constraint rules for specific vulnerability types, allowing for real-time suspension of flawed contract transactions at the Ethereum Virtual Machine (EVM) level, even post-deployment. They addressed the limitations of traditional expert rule-based and deep learning approaches by combining them, improving interpretability and accuracy in vulnerability detection. Using this two-phase mechanism, they demonstrated a significant improvement in vulnerability detection effectiveness, achieving an average score of approximately six points higher than the GNN model alone. Empirical results indicated that the mechanism not only enhanced detection but can also block harmful contract transactions at the EVM level and generate detailed error reports, contributing to a more secure smart contract environment.

\cite{han_smart_2022} utilised GNN to improve semantic and structural features captured in smart contract source code. They incorporated CNNs for learning node order within the control flow graphs (CFG) and combined GNN and CNN in a single model for enhanced vulnerability detection. They validated their approach with real Ethereum smart contract datasets, showing increased accuracy and F1 values. In their later work, \cite{han_smart_2023} introduces a deep learning-based vulnerability detection model that fuses syntactic and semantic features of smart contract source code for improved vulnerability detection. They address the limitations of current models that focus on single representation forms of source code, which are insufficient for catering rich semantic and structural information. The framework employs TextCNN and GNNs to extract features from Abstract Syntax Trees (AST) and Control Flow Graphs (CFG) of smart contracts. The framework achieves high detection accuracy recall rates on detecting five types of smart contract vulnerabilities, boasting an average precision of 96\% and a recall rate of 90\%. The framework also demonstrates that feature fusion vectors offer better characterisation of vulnerability-related features in source code, leading to effective and rapid detection of vulnerabilities.

\cite{cai_combine_2023}'s GNN approach constructs a graph representation of smart contract functions, incorporating syntactic and semantic features by merging AST, CFG and Program Dependency Graphs (PDG). They utilise program slicing techniques to normalise the constructed graph and remove redundant information, making it more focused on vulnerability-related features. They also introduced a Bidirectional Gated Graph Neural Network model with a hybrid attention pooling layer to capture vulnerability-related code features effectively. Their work achieved high empirical performance with 89.2\% precision and 92.9\% recall on a dataset of nine common types of vulnerabilities and small contracts. The paper also demonstrates superior performance compared to existing state-of-the-art vulnerability detection approaches.
\cite{liu_combining_2021} fuses expert patterns and GNNs. They introduce and normalise "contract graphs" to highlight key code elements and propose a novel "temporal message propagation network" for feature extraction, outperforming existing, rule-based methods in detecting three vulnerability types. 

\cite{zeng_ethergis_2022}'s EtherGIS model is a scalable framework for multi-type and large-scale vulnerability detection. The paper proposes an algorithm that converts smart contracts into attribute graphs, capturing contract behaviour and specific vulnerability traits. It incorporates Graph Neural Networks (GNN) to synthesise low-level graph features into high-level contract features and constructs a robust dataset for performance assessment, demonstrating superior accuracy and stability compared to existing approaches.

\cite{nguyen_mando_guru_2022} introduces MANDOGURU, a deep learning tool employing control-flow and call graphs with graph attention neural networks. The model detects vulnerabilities at contract and line levels and improves the F1-score by up to 24\% at the contract level, 63.4\% over traditional methods. In another paper, \cite{nguyen_mando-hgt_2023} proposes MANDO-HGT, which employs heterogeneous contract graphs (HCGs) to represent control-flow and function-call information in smart contracts. MANDO-HGT significantly improved detection accuracy over state-of-the-art methods, with F1-score gains ranging from 0.7\% to 76\%. MANDO-HGT also presents flexibility in being retrained for different types of vulnerabilities without requiring manually defined patterns. A further addition to their MANDO framework sees \cite{nguyen_mando_2022} proposes a graph representation for Ethereum smart contracts based on heterogeneous control flow and call graphs. MANDO utilises customised metapaths and a multi-metapath attention network for nuanced vulnerability detection and achieves significant F1-score improvements in fine-grained and coarse-grained vulnerability detection.

\cite{fan_smart_2021} introduces a Dual Attention Graph Convolutional Network (DA-GCN) for smart contract vulnerability detection, which employs a GCN for feature extraction from generated CFGs. Incorporating a multilayer perceptron (MLP) for final vulnerability classification, the model achieved high accuracy rates of 91.2\% and 87.5\% in detecting the two types of vulnerabilities. 
\cite{chen_smart_2023} introduces a GNN model specifically for detecting vulnerabilities in smart contracts within educational blockchains. They demonstrate that bytecode files serve as an effective source for vulnerability detection and establish that the model's predictive accuracy improves with the addition of semantic processing and edge classification. The model outperforms traditional methods and achieves good results with fewer layers in the GCN.
\cite{zhuang_smart_2021} proposes a degree-free graph convolutional neural network (DR-GCN) and a novel temporal message propagation network (TMP) to learn from normalised graphs developed to represent smart contract code.

\cite{liu_smart_2021} investigated combining vulnerability-specific expert patterns with neural networks in an interpretable manner, using a simple yet effective multi-encoder network for feature fusion. They create a semantic graph to represent code flow, adopt a TMP GNN for global graph transformation, and fuse expert patterns and graph features via an attentive multi-encoder network. They find that combining expert patterns with graph features is crucial for detection performance, as demonstrated by consistently better results than variants. 
\cite{zhang_toward_2022} used an Abstract Semantic Graph (ASG) to capture syntactic and semantic features in smart contracts and applied a GNN and Graph Matching Network for vulnerability detection. This method outperformed existing tools, improving F1 scores by 12.6\% for source code and 25.6\% for bytecode. \cite{luo_vuldet_2023} also constructs a contract graph by treating all variables and function calls in a smart contract as nodes and capturing their control flow and data flow information. They employ Graph Attention Networks (GAT) to analyse the contract graph and utilise a multi-head attention mechanism to stabilise the learning process during GAT application. They demonstrate through extensive experiments that VULDET excels in identifying reentrancy and timestamp dependency vulnerabilities.

\cite{liu_vulnerable_2023} introduces a Multi-Relational Nested contract Graph (MRNG) for comprehensive characterisation of smart contract code, including relationships between functions. They create Multi-Relational Function Graphs (MRFGs) to represent individual function code within the MRNG, and propose a Multi-Relational Nested Graph Convolutional Network (MRN-GCN) to extract semantic and structural features from MRFGs and MRNG. MRN-GCN for vulnerable smart contract functions enhances accuracy, precision, recall, and F1-score.

\subsubsection{Hybrid Models}

\cite{xu_hybrid_2023} proposes a hybrid CNN-GRU neural network model, utilising a CNN for local feature extraction and a GRU for handling sequential data and dependencies. The model achieves 94\% average precision in detecting six types of vulnerabilities and outperforms base models as well as traditional ML models such as SVM and RF in accuracy metrics. \cite{jie_novel_2023} proposes a methodology for smart contract vulnerability detection that leverages strong white-box knowledge in a series of supervised multimodal tasks under static analysis. The approach leverages NLP techniques and utilises multiple features, such as code and graph embeddings, in both intramodal and intermodal settings. It also employs state-of-the-art machine learning algorithms like self-attentive bi-LSTM, textCNN, and Random Forest (RF) for the tasks. An extensive evaluation is performed on the SmartEmbed dataset with 101,082 functions, revealing that the framework outperforms existing schemes. The highest detection performance achieved was 99.71\%. \cite{zhang_novel_2023} introduces BiGAS, a composite model comprised of BiGRU, attention mechanisms and SVM for reentrancy vulnerability detection. They achieved accuracy and F1-score above 93\% in detecting reentrancy vulnerabilities, outperforming existing state-of-the-art smart contract vulnerability detection tools, and demonstrated that replacing Softmax with SVM as the classifier improved model performance, achieving an accuracy of 93.24\% and an F1-score of 93.17\%. Comparative analysis revealed that the BiGAS model improved accuracy and F1-scores by 4 to 23\% over existing advanced automated audit tools and deep learning-based methods. 

\cite{zhang_cbgru_2022} introduces a hybrid deep learning model, CBGRU, fusing multiple word embedding techniques (Word2Vec, FastTest) and deep learning architectures ((LSTM, GRU, BiLSTM, CNN, BiGRU) for vulnerability detection. They empirically demonstrate that CBGRU outperforms existing single neural network models in detecting vulnerabilities on the SmartBugs Dataset-Wild, achieivn over 90\% accuracy and F1 scores 4 out of 6 vulnerabilities. \cite{jie_full-stack_2021}’s model employs a full-stack feature extraction approach, integrating source code, build-based, and EVM bytecode features for vulnerability detection. They employ a BiLSTM with self-attention mechanism for learning and employ TexctCNN, RF, GCN and CNN for feature extraction. They found that using the SMOTE strategy with Word2Vec embedding seems to offer the best overall performance among most metrics. \cite{li_reentrancy_2023} introduces a novel reentrancy vulnerability detection method that leverages Conv1DBiGRU and expert knowledge. They combine convolutional models, sequence models, and traditional expert knowledge, achieving superior performance in accuracy, recall, precision, and F1 score, with the F1 score reaching 90.71\%, outperforming existing deep learning-based and traditional methods.

\cite{wu_smart_2023} introduces a Hybrid Attention Mechanism (HAM), improving upon the limitations of traditional and deep learning models. They extract code fragments that concentrate on critical points of vulnerability and fuse single-head and multi-head attention features specifically for enhancing detection. They report specific detection accuracies ranging from 80.85\% to 93.36\% for five types of vulnerabilities. \cite{zhang_spcbig-ec_2022} introduces a novel hybrid model called Serial-Parallel Convolutional Bidirectional Gated Recurrent Network Model incorporating Ensemble Classifiers (SPCBIG-EC). They developed a unique CNN structure, Serial-Parallel CNN, to effectively capture both local and global features of the smart contract code sequences and utilised an ensemble classifier in the model to bolster its robustness by aggregating the decisions of multiple classifier experts. SPCBIG-EC outperformed 11 other advanced methods, achieving high F1-scores for detecting reentrancy (96.74\%), timestamp dependency (91.62\%), and infinite loop vulnerabilities (95.00\%). \cite{zhang_svscanner_2023}’s SVSCanner method combines global semantic features and deep structural semantics from AST, and utilises a TextCNN as the classifier. Deep semantic extraction includes global feature extraction using Bi-LSTM and structural feature extraction using a multi-head attention mechanism. Experimental results show a 7.33\% accuracy improvement compared to traditional methods. \cite{xu_vulnerability_2023} introduces SolBERT-BiGRU-Attention, which combines a lightweight pre-trained BERT model with a Bidirectional Gated Recurrent Unit (BiGRU) neural network incorporating a hierarchical attention mechanism, demonstrating improved detection accuracy, achieving a 93.85\% accuracy rate and a 94.02\% Micro-F1 Score in identifying vulnerabilities in smart contracts. 

\cite{duy_vulnsense_2023} produces VulnSense, a multimodal learning framework incorporating BERT, BiLSTM, and GNN models to analyse multiple types of features—source code, opcode sequences, and control flow graphs (CFG). They leverage three different datasets (Curated, SolidiFI-Benchmark, and Smartbugs Wild) for comprehensive feature fusion, enhancing the model's capability to capture semantic relationships during analysis. They achieve an average accuracy of 77.96\% in detecting vulnerabilities across three categories of smart contracts. \cite{xu_w2v-sa_2023} introduces W2V-SA, combining a CNN with a self-attention mechanism. Experimental results demonstrate W2V-SA outperforms standalone neural network models and other machine learning techniques in vulnerability detection, achieving an average accuracy of over 94\% targeting six vulnerabilities. \cite{wang_detecting_2023} employs a CNN-BiLSTM architecture to analyse opcode sequences in transactions. The model is capable of dynamic detection of vulnerabilities even after the smart contract is live, offering the potential to minimise losses. This model was extensively evaluated on over 1500 real-world opcode sequences and achieved an accuracy of 82.86\% and an F1-score of 83.63\%.

\subsection{Other Machine Learning Techniques}

\subsubsection{Natural Language Processing}

\cite{duan_new_2023} introduces a novel smart contract vulnerability detection approach that fuses features extracted from both opcodes and source code. The model uses 2-gram features from opcodes and token features from source code via a pre-trained CodeBERT model, capturing semantics at multiple levels. The model achieves high detection accuracy rates—98\% for reentrancy and timestamp dependence and 94\% for transaction-ordering dependence.
\cite{sun_assbert_2023} proposes ASSBert, a framework designed for smart contract vulnerability classification with limited labelled data. They merge active and semi-supervised learning into BERT to select valuable code data from unlabeled sets. ASSBert outperformed baseline models Bert, Bert-AL, and Bert-SSL in vulnerability detection across multiple metrics and varying levels of labeled data.  \cite{wu_peculiar_2021} introduced Peculiar, a novel approach for smart contract representation based on crucial data flow information to capture key features without overfitting. They note that using a Crucial Data Flow Graph (CDFG) in Peculiar is essential for performance. When streamlining is removed (Peculiar-WOS variant), there is a substantial drop in both recall (35.36\%) and F1 score (23.35\%). \cite{zhang_reentrancy_2022} posits ReVulDL, a deep learning-based two-phase debugger for smart contract reentrancy vulnerabilities that leverages GraphCodeBert. The model employs a two-phase approach, firstly undergoing detection using a graph-based pre-training model to understand complex relationships in propagation chains and then undergoing localisation through interpretable machine learning to pinpoint suspicious statements if a vulnerability is detected. The model outperforms 16 state-of-the-art vulnerability detection approaches for reentrancy vulnerabilities. \cite{jeon_smartcondetect_2021} proposes SmartConDetect, a BERT-based model for detecting vulnerabilities. SmartConDetect achieves a higher F1-score than other baselines, including SVM, Eth2Vec, and GNN, demonstrating its effectiveness in vulnerability detection. \cite{jiang_vddl_2023} proposes the VDDL model, which is a multi-layer bidirectional Transformer architecture with multi-head attention and masking mechanisms combined with CodeBERT to improve training. VDDL achieved an accuracy of 92.35\%, recall of 81.43\%, and F1-score of 86.38\% for efficient vulnerability detection in smart contracts.

\subsubsection{Other Techniques}

\cite{zhang_novel_2022} introduces SCVDIE, an ensemble learning model for smart contract vulnerability prediction using seven pre-trained neural networks based on an Information Graph (IG). The model outperformed static tools and seven other data-driven methods in accuracy and robustness. \cite{narayana_automation_2023} addresses a deep learning model for enhanced detection of re-entrancy, DOS, and Transaction Origin attacks. They employ binary, multi-class, and multi-label classification techniques for improved results.  \cite{gong_gratdet_2023} proposes GRATDet, a novel vulnerability detection approach for smart contracts that integrates deep learning, graph representation, and Transformer models. They developed a data augmentation technique to address class imbalance by performing syntactic analysis and function combinations in Solidity source code. They utilised an improved Transformer–GP model that jointly learns node and edge information from the LG, fusing global and local features for precise vulnerability detection. The model demonstrated superior performance in reentrancy vulnerability detection, achieving an F1 score of 95.16\%, outperforming existing state-of-the-art methods. \cite{zhang_smart_2022-1} introduces aMultiple-Objective Detection Neural Network (MODNN) for scalable smart contract vulnerability detection, validating 12 types of vulnerabilities, including 10 known and additional unknown types, without requiring specialist knowledge. The model achieves a 94.8\% average F1 Score in tests, outperforming existing machine learning models for this application.
\cite{huang_smart_2022} introduces a multi-task learning model for smart contract vulnerability detection. The model architecture consists of two parts: 1) A bottom sharing layer using attention-based neural networks to learn semantic information and extract feature vectors; 2) A task-specific layer using CNNs for classification tasks, inheriting features from the shared layer. \cite{guo_smart_2022} develops a Siamese network model called SCVSN for source code-level vulnerability detection. Their model outperforms previous deep learning-based methods in experiments. \cite{zhang_vulnerability_2022} introduces BwdBAL to tackle the problem of lacking labelled data, in which they utilise Bayesian Active Learning (BAL) to minimise model uncertainty during the sampling phase in active learning. They demonstrate that BwdBAL outperforms two baseline methods in smart contract vulnerability detection and confirm that the uncertainty sampling strategy is superior to four other sampling strategies.

\subsection{Summary of Trends}

We now distil the essential methodologies and outcomes from the literature we have explored, highlighting the trends across model usage in their application to smart contract vulnerability detection.

\begin{itemize}
    \item Classical machine learning techniques emphasise the usage of models like KNN, RF, DT, XGBoost and SVM, which outperform static tools. 
    \item Multi-model approaches tend towards integrating complex feature extraction methods and optimisation techniques using a combination of deep learning and classical machine learning approaches. Techniques like slicing, control-flow graphs and feature extraction from bytecode have been introduced and show improvement in precision and recall, revealing significant improvements over earlier models. Further, fuzz testing methods synergised with machine learning have been shown to assess vulnerabilities with high accuracy.
    \item CNNs often leverage abstract syntax trees, vector space mappings, and advanced feature engineering, including self-attention mechanisms, to enhance detection capabilities. Techniques range from translating bytecode into RGB colour codes for image-based CNN analysis to employing critical combination paths for focusing on code sections with potential vulnerabilities. Experimental results demonstrate that CNN-leveraged techniques surpass traditional benchmarks in detection accuracy and computational efficiency.
    \item Research employing RNNs demonstrates models like LSTM, GRU and BiLSTM outperform traditional models and symbolic analysis tools in accuracy, precision, recall and time efficiency.
    \item GNNs have been employed to capture the semantic and structural features of smart contracts. Graph representations such as Control flow Graphs and Program Dependency Graphs are leveraged to analyse the code effectively. Studies also focus on techniques geared towards the refining of the feature space for GNNs to improve precision and recall rates.
    \item Hybrid machine learning models are at the forefront of vulnerability detection, adapting and combining different neural network architectures to boost improvements in precision and recall. Employing a wide array of techniques, like CNN for feature extraction and GRUs for temporal dependency learning, several hybrid models have achieved near-perfect performance metrics, representing a significant advancement in the improvement of accuracy in this discipline.
    \item Although it represents a small segment of currently available techniques, NLP-based approaches have leveraged pre-trained models like CodeBERT and BERT to survey opcode and source code semantics, capturing multi-level semantic features and combining token features and code structures for improved accuracy. Data flow graphs with NLP models have further refined the detection process as well. Some of these methods achieve almost perfect detection rates, too.
    \item Other novel techniques  range from ensemble models to transformers, some of which aim to handle class imbalances and address the challenge of limited labelled data.
   
\end{itemize}

\section{Discussion} \label{chap:discussion}

In this section, we describe the critiques, limitations and future work for each of our proposed papers, outlining where further experimentation should take place in future research. In so doing, we address our final research question \textbf{(RQ3)}: What are the current research gaps and future work opportunities in the application of machine learning for enhancing the detection and mitigation of vulnerabilities in smart contracts? We aim to elucidate the existing research gaps and future opportunities to offer insights that can inform the design of future endeavours.

\subsection{Traditional Machine Learning Techniques}

\cite{mandloi_machine_2022} relies on pre-labelled transaction data for training, which limits real-world applicability. The authors suggest generating labelled transactions dynamically and establishing both benign and malignant user contracts, as well as incorporating multiple types of machine learning algorithms in future work. \cite{xu_novel_2021} notes that their model does not detect concrete problems and cannot locate the line of the code where the vulnerability occurs in the smart contract. They mention additional data collection for a more comprehensive set of basic malicious smart contracts is also planned. \cite{yang_smart_2022} plans on refining AST-based feature extraction by selecting more representative information in different types of AST nodes for future work. They also aim to incorporate deep learning techniques, such as LSTM and GRU, to potentially improve vulnerability performance. While \cite{zhao_novel_2023} proposes an automatic labelling system, it’s unclear how accurate this is compared to manual labelling. The paper only focuses on reentrancy vulnerabilities and emphasises empirical results, but lacks any theoretical backing as to why the method works as well as it does. \cite{song_efficient_2019} is dependent on bigram features extracted from simplified opcodes, representing only the static characteristics of smart contracts. Future investigations will look into designing models capable of detecting novel vulnerabilities, extending the model’s capabilities.
\cite{wang_contractward_2021} is also dependent on bigram features, and is dependent on Oyente for labels, which could introduce bias or errors. Future work includes exploring more effective features to improve performance. \cite{li_detecting_2023} plans on employing trigram models for better feature extraction in the future, and intend to validate their scheme using a broader range of machine learning models. In future work, the paper would benefit from the delineation of all results, as accuracy and f1-scores were only presented pictorially.  \cite{eshghie_dynamic_2021} suggested the need for combining machine learning-based detection with oracle-supported dynamic vulnerability detection to reduce false negatives and emphasises the importance of randomness in generated transactions for detecting complex attacks. They also wish to investigate automatic test-case generation tools and the addition of more features for more accurate detection.  
\cite{momeni_machine_2019} wishes to involve training on more diverse datasets, incorporate more static analysers, and extend feature sets for higher accuracy. \cite{hao_scscan_2020} wishes to expand the sample space and incorporate deep learning for more effective code analysis, as well as extend the system for scanning vulnerabilities in other types of smart contracts, not just Ethereum-based contracts. \cite{lakshminarayana_towards_2022}’s conversion tool, BPMN-SOL, only generates a template, and requires manual effort to create the final smart contract. For future work, they should focus on improving the smart contract template quality, and aim for a fully automated smart contract conversion, as well as potentially explore more advanced deep learning techniques to further enhance vulnerability detection accuracy. \cite{lohith_tp-detect_2023} wishes to explore deep learning and transfer learning methods on the TriPix dataset, and employ the adaptive nature of the model architecture by extending it to detect future vulnerabilities.

\subsection{Neural Networks and Deep Learning}

\subsubsection{General Multi-Model Usage}

\cite{xing_new_2020} plans to better integrate slicing matrix features with RF models to possibly further improve vulnerability detection, and explore other detection methods to fully tap into the potential of the slicing matrix feature technique. \cite{yu_deescvhunter_2021}’s work is limited to reentrancy and time-dependence vulnerabilities. Real-world scenario testing and further evaluation may provide a more comprehensive assessment. \cite{wu_detecting_2022}’s work relies solely on static opcode features, which may pose as a limitation. Future work includes exploring novel features and incorporating function dependencies. \cite{lutz_escort_2021}’s performance could be improved on specific vulnerability types, like the “self destruct” vulnerability, which bore a lower recall score. The effectiveness of transfer learning could vary with the introduction of more complex vulnerability types, which should be studied further. \cite{ashizawa_eth2vec_2021} operates in an unsupervised setting; exploration of supervised settings could improve vulnerability classification. There are no inter-contract analysis capabilities; therefore, the framework will miss vulnerabilities involving multiple contracts. \cite{li_smart_2022} could look into further feature construction work. \cite{deng_smart_2023} should explore using small samples and unsupervised learning to mitigate data imbalance issues. They also call on the need for a standard, unified smart contract vulnerability dataset for more reliable experimentation and comparative analysis. \cite{sendner_smarter_2023} is limited to specific vulnerabilities, and its performance may vary with changing coding styles and opcode mappings over time.  \cite{yuan_svchecker_2022} relies on Oyente for labelling, which has its own limitations. The work could benefit from incorporating additional features and exploring the tuning of parameters. \cite{mi_vscl_2021} is limited to analysing smart contracts in binary representation. \cite{gopali_vulnerability_2022}’s “normal” classification is not foolproof and could still contain unidentified vulnerabilities, which may lead to a slight skewing of the results. \cite{xue_xfuzz_2022}’s work may still miss vulnerabilities with complex path conditions and doesn't support some types of vulnerabilities. They wish to improve the handling of complex path conditions to avoid missing vulnerabilities and extend xFuzz with static analysis to detect more types of vulnerabilities. \cite{liao_soliaudit_2019} plans to test dynamic opcode sequences and improve input generation with symbolic execution.

\subsubsection{Convolutional Neural Networks}

\cite{zhou_vulnerability_2022} declares the need for graph-based convolution for nonrectangular structured data; currently, there is no unified approach to generating structure data like CFGs, which could affect reliability. They also suggest that the rapid evolution in smart contract standards poses challenges for identifying new vulnerabilities, pointing to a need for improved unsupervised methods. 

\cite{xu_efficient_2023} suggests improving the vector representation for broader and more accurate vulnerability detection in the future. Further experimentation is recommended surrounding the code obfuscation component of the methodology, as obfuscation may introduce new variables in detection, which should be taken into account when considering the effectiveness of the proposed method. \cite{sun_attention-based_2021}’s model is limited to detecting three types of vulnerabilities, and functions primarily as a pre-filter for more traditional symbolic analysis tools rather than to serve as a comprehensive solution. \cite{sun_attention-based_2021} intends to extend the model from binary to multinomial classification for detecting multiple vulnerabilities simultaneously and plan to integrate the proposed method with formal verification techniques to improve the model’s accuracy further. While \cite{hwang_codenet_2022}’s CodeNet model excelled in accuracy and recall, its precision was slightly lower than some state-of-the-art models. The paper offers limited insights into its limitations or weaknesses and does not address the false positive rate in depth. A more extensive dataset for evaluation incorporating more vulnerability types could offer a deeper understanding of CodeNet’s effectiveness. \cite{huang_hunting_2018} aims to optimise the parameters and network structure of their approach. The proposed system presents high accuracy but does not expound upon the labor-cost evaluation considering its practical focus. \cite{rossini_smart_2022} declares the potential of imbalanced class distribution bearing the potential for affected model performance, and that the dataset they employ may not capture all smart contract vulnerabilities. They also address challenges posed by evolving smart contract vulnerabilities, instigating the necessity for adaptable models and newer datasets.  \cite{qin_smart_2022}’s work does not mention false positive or negative rates, which are critical for evaluating the effectiveness of a vulnerability detection scheme, and only one other deep learning vulnerability detection scheme is considered for performance comparison.

\subsubsection{Recurrent Neural Networks}

\cite{gupta_deep_2022}’s penalisation side of the reward system could deter benign activity if false positives occur. Further, their LSTM model has not been tested against various vulnerabilities beyond the dataset, leaving its applicability uncertain. \cite{wang_ethereum_2023}’s model can only detect the presence of vulnerabilities, not the type of them. Their claim of enhanced interpretability is not empirically substantiated, and there is a limited comparative analysis with state-of-the-art methods. They aim to develop a multi-label classification for specific vulnerabilities in the future and integrate other machine learning techniques for improved detection. \cite{gogineni_multi-class_2020} addresses the challenge of generating more accurately classified smart contracts to improve the model and that refinement in training data and further exploration of certain vulnerability types could enhance the model’s efficacy. \cite{qian_multi-label_2022} notes overfitting issues in their RNN and lSTM models, affecting performance on less common vulnerabilities. In future work, they recommend identifying operations leading to vulnerabilities and using end-to-end NLP technologies for opcode processing. \cite{xu_reentrancy_2022} proposes trying more innovative methods to detect other types of smart contract vulnerabilities. It is recommended they draw more comparisons with traditional and newer machine learning-based techniques for analysis. \cite{ren_smart_2023}’s Blass approach only considers four types of vulnerabilities. Future work aims to support binary code and include more types of vulnerabilities for detection. \cite{zhang_smart_2022}’s recall performance is suboptimal, affecting F1 scores across different vulnerabilities. The paper’s methodology has room for improvement, particularly in the area of recall metrics, and future work could focus on enhancing recall and the robustness of the model. \cite{zhou_smart_2023} aims to explore new architectures for effective task handling and feature representation and plan on collecting more data on vulnerable contracts for improved training performance. \cite{hu_smart-contract_2023}’s work is limited to four types of vulnerabilities - future work could involve extending the dataset to include a broader range of vulnerabilities for more comprehensive testing. Comparative analysis with other methodologies could strengthen the assertion that the model has good coverage and efficiency. \cite{goswami_tokencheck_2021} does not compare with other machine learning or deep learning models to justify the choice of LSTM. It also does not address generalisability, given the domain is strictly ERC-20 contracts. Actual deployment and integration challenges are not discussed. \cite{qian_towards_2020} declares their plans to extend the model to detect other types of vulnerabilities like integer overflow and unhandled exceptions. They also aim to design scenarios for runtime state analysis to evaluate dynamic execution vulnerabilities. \cite{tann_towards_2018} holds limitations in control-flow analysis in that LSTM struggles with complex control-flow properties like loops and internal function calls. Future work includes exploring other state-of-the-art models and addressing the challenge of obtaining more accurate labels for improved training performance. \cite{qian_bilstm-attention_2022} lacks traditional model comparison and does not detail the selection process for their hyperparameter tuning. Future work includes temporal analyses; studying model adaptability over time and combining with other models for better accuracy.

\subsubsection{Graph Neural Networks}

\cite{liu_smart_2023} proposes enhancing the efficiency of GNN’s aggregation algorithm and improving the generation of contract graphs to better capture contract code semantics. They also wish to minimise system runtime overhead by optimising the spatial structure of core control flow and opcode records. \cite{han_smart_2022} notes that they wish to expand the current model for a more nuanced representation of CFGs and that while the model shows effectiveness on real vulnerability datasets, it leaves room for incorporating additional types of semantic and control dependency information. Future work in \cite{han_smart_2023} aims to develop models that can pinpoint the exact location of vulnerabilities within the code. Further, the model currently doesn’t define complex vulnerability patterns, suggesting that there may be room for more complex detection strategies in future iterations. The dataset in \cite{cai_combine_2023}’s work may be prone to interrogation, given it was labelled by humans. \cite{cai_combine_2023} also declares sample imbalances and hyperparameter issues and say they want to include more graph types in the future. \cite{liu_combining_2021} only focuses on function-level vulnerabilities, not the entire smart contract. The method wasn’t tested on smart contracts that only have bytecode. \cite{liu_combining_2021} plans on exploring the architecture’s effectiveness on a broader range of vulnerabilities in the future. \cite{zeng_ethergis_2022}’s framework is heavily dependent on expert knowledge to analyse EVM bytecode for identifying potential vulnerabilities. Although the model is superior to existing methods, EthergIS has room for improvement in reducing false negatives for certain vulnerabilities. Further, another area of research to investigate may be in the design of a specialised neural network to locate specific graph areas where vulnerabilities exist.


Given the novelty of \cite{nguyen_mando_guru_2022}’s technique, comparisons drawn with established methods may be limited. Future work may involve expanding the types of vulnerabilities detected and improving generalisability. \cite{nguyen_mando-hgt_2023} depends on external analysis tools for generating control flow and call graphs. It lacks consideration for certain bug types with less than 50 contracts in the dataset. Therefore, future endeavours would benefit from expanding the dataset to cover a wider range of vulnerabilities. Embedding techniques can be improved to include more semantic properties like data dependencies in \cite{nguyen_mando_2022}’s following paper. \cite{fan_smart_2021}’s paper is limited to reentrancy and timestamp dependency vulnerabilities. It does not investigate real-time detection or proactive vulnerability prevention and does not discuss false positives or negatives in-depth.  \cite{chen_smart_2023}’s work should incorporate additional types of vulnerabilities for a more comprehensive approach in the future. The lack of comparison with existing vulnerability detection tools may limit the method’s generaliability.  \cite{10050059} paper requires that the network depth must be carefully chosen, and notes that blindly increasing hidden layers is not advised. Future work could also focus on generating better graph structures from either smart contract source code or bytecode. \cite{zhuang_smart_2021} outperforms existing methods and other neural networks, however, the system’s robustness against new types of vulnerabilities remains untested. \cite{liu_smart_2021} presents a limited focus on discussing potential limitations or critiques of their approach. However, further research into the interpretability of weight distributions and evaluation of a broader range of vulnerabilities could be performed. \cite{zhang_toward_2022} notes that they wish to expand their research to cover more types of smart contract vulnerabilities to improve scalability and generalise the approach for application in other software testing scenarios (different programming languages and code clone detection). \cite{luo_vuldet_2023} lacks a comparative analysis against non-neural network-based approaches and bears an absence of the discussion of potential false positives and negatives plus adaptability to evolving vulnerabilities.  \cite{liu_vulnerable_2023} mentions involving expert knowledge and semi-supervised learning to address dataset reliability concerns in the future, and note hyperparameters and model initialisation randomness as internal threats to the validity of the paper. 

\subsubsection{Hybrid Models}

\cite{xu_hybrid_2023}’s model requires high computational time and resources for training and detection, calling for future optimisation. Their model is also more effective in detecting vulnerabilities with prominent features and needs adaptation for those with complex or subtle features. \cite{jie_novel_2023} notes that its reliance on word2vec results in a lack of support for out-of-vocabulary words, and suggests fastText as a solution. They also call for future work on multi-class classification. Further, while the framework can analyse contracts without source code, it indicates that relying solely on bytecode (EVMB layer) leads to poor detection performance. \cite{zhang_novel_2023}’s model is limited to detecting reentrancy vulnerabilities only, with future work aiming to expand the types of vulnerabilities detected. They also plan to design a plausibility verification method to cross-check automated and manual detection results and suggest that a unified model combining deep learning and expert rules is a key future research direction. \cite{zhang_cbgru_2022} reports high accuracy, but only reports on three vulnerabilities, raising questions about generalisability. The paper also does not delve into computational cost, scalability, or real-world deployment concerns. \cite{jie_full-stack_2021}’s work excludes expert patterns, possibly limiting detection and is limited to binary classification of vulnerabilities. They also make unvalidated claims of resilience to missing data. \cite{li_reentrancy_2023} proposes building a more comprehensive and larger dataset for smart contract vulnerability detection research. \cite{wu_smart_2023}’s model is a black box, lacking transparency in how it determines vulnerabilities. The model was only tested against specific vulnerabilities and may not generalise well to untested types. Exploring other embedding techniques could also yield different results. A lack of investigation into the computational complexity and resource requirements could hinder its translation to real-world applicability. \cite{zhang_spcbig-ec_2022} also notes a need for a unified and standardised smart contract vulnerability dataset to further improve deep-learning-based detection methods. \cite{zhang_svscanner_2023} mentions investigating data dependency information between critical variables for richer semantic information and improved accuracy in future work.  \cite{xu_vulnerability_2023} wishes to expand to more types of smart contracts for improved model robustness and incorporate more lightweight pre-trained language models to reduce training costs and improve detection accuracy. \cite{duy_vulnsense_2023} does not mention any explicit limitations or future work, suggesting a potential gap in the study’s threats to validity. \cite{xu_w2v-sa_2023} wishes to study vector representation and feature extraction methods to enhance efficiency further and plan to improve model structure through the use of sentence vectors or other word embedding models. They also wish to integrate different attention mechanisms to refine the approach.  \cite{wang_detecting_2023} plans to incorporate attention mechanisms into the model to potentially improve its performance.

\subsection{Other Machine Learning Techniques}

\subsubsection{Natural Language Processing}

\cite{duan_new_2023} aims to improve generalisability by collecting more data on different vulnerability types and employing techniques like generative adversarial networks for adversarial training. \cite{sun_assbert_2023} may integrate stream-based active learning methods to reduce program execution time and operational costs and consider data distribution alongside representativeness in that they plan to build a density estimator to refine the sampling strategy. \cite{wu_peculiar_2021}’s paper does not address how Peculiar could scale with larger or more complex smart contracts, leaving questions regarding real-world scenarios. The paper also hints at the idea of using pre-trained models for detection. \cite{zhang_reentrancy_2022} mentions possible bugs in baselines despite implementation and testing and that experimental results might not apply universally due to complex factors in real-life development. Future work suggests enhancing interpretable machine learning and extending the approach to detect and locate other vulnerabilities. \cite{jeon_smartcondetect_2021} intends to develop a more advanced model for detecting vulnerabilities using few-shot learning. \cite{jiang_vddl_2023}’s work is limited to source code and could benefit from using bytecode-only contracts in training. 

\subsubsection{Other Techniques}

\cite{zhang_novel_2022}’s use of only binary classification limits the significance of the results, future work will aim to categorise different types of vulnerabilities.  \cite{narayana_automation_2023}’s model is limited to three types of vulnerabilities. Future work will explore additional deep learning models not covered in the study. \cite{gong_gratdet_2023}’s paper only focuses solely on reentrancy vulnerabilities, and future work aims to extend the dataset to include other vulnerability types. While the paper demonstrates robustness in vulnerability detection, the claim is somewhat constrained by the limited scope of vulnerabilities studied. \cite{wang_gvd-net_2022}’s paper lacks an empirical comparison with manual methods when it claims it performs better than manual methods. \cite{zhang_smart_2022-1} relies on opcodes and does not fully consider syntactic and semantic information of the original smart contract code, limiting its feature extraction capabilities. \cite{huang_smart_2022} outperforms existing tools and machine learning methods in precision, but limitations include hard parameter sharing and manual weight settings for the loss function. \cite{guo_smart_2022}’s model shows excellent performance but is limited to detecting reentrancy vulnerabilities, future work points to detecting more types. \cite{zhang_vulnerability_2022} considers adopting diverse feature extraction methods for different vulnerabilities to improve precision. 

\subsection{Summary of Recommendations for Future Work} \label{sec:futureWork}

We now distil the essential recommendations for future work, emphasising the most prominent critiques and limitations across the research.

\begin{itemize}
    \item In the realm of CNNs, future work should focus on developing convolutional techniques tailored to CFGs, improving vector representations, extending models to multinomial classification, expanding datasets to better evaluate models, addressing imbalanced class distributions, and thoroughly evaluating false positive and negative rates.
    \item For more general multi-model usage, future work should focus on expanding vulnerability detection beyond reentrancy and time-dependency vulnerabilities, enhancing static and dynamic analysis features, developing methods to mitigate data imbalance, creating a standard, universal vulnerability dataset and investigating models that are adaptable to evolving smart contract coding conventions.
    \item Future work in GNNs should focus on expanding pattern recognition to include complex dependencies and location pinpointing of vulnerabilities, investigating hyperparameter optimisation, broadening the scope of vulnerabilities studied and advancing the robustness and generalisability of current detection methods to combat evolving vulnerabilities.
    \item For hybrid models, optimising computational efficiency, ensuring model transparency, standardising datasets and incorporating even more advanced machine learning mechanisms could be explored.
    \item Parties interested in employing NLP for vulnerability detection should focus on the improvement of data diversity and model generalisability, exploring pre-trained models for scalability, and extending detection capabilities to bytecode analysis and few-shot learning techniques.
    \item For other novel techniques, generally, incorporating a wider range of vulnerabilities and deep learning models would be beneficial. 
    \item RNN research should tackle model generalisability, enhancements to multi-label classification, reduction in overfitting for less common vulnerability types, and comparative analyses against a wider range of machine learning techniques. Temporal analyses confirming the effectiveness of model adaptability could also be of interest.
    \item Future efforts regarding classical machine learning techniques should focus on combining these techniques with dynamic analysis and exploring transfer learning to enable the adaptability of detecting vulnerabilities as they evolve.
    
\end{itemize}


\section{Conclusion} \label{sec:conclusion}

This article has proffered an extensive, systematic review of the ways in which machine learning has the ability to detect and mitigate vulnerabilities within smart contracts. The research addresses the breadth of machine learning methodologies currently being experimented with to identify and counteract specific vulnerabilities.  Our synthesis and analysis of the state-of-the-art and the research landscape has illuminated the efficacy and limitations of frameworks from classical machine learning algorithms to more complex neural network architectures in attacking this specific problem. Furthermore, this study has expounded the current research gaps and revealed potential avenues for further research, guiding successive empirical investigations in this domain. Overarchingly, through the work's rigorous methodological framework, we have also provided a seminal repository, which shall serve as a practical resource for parties invested in blockchain security. Such a resource could prove essential for efficient navigation of the current terrain of knowledge to drive further empirical investigations and inform practitioners in their investigations. Machine learning holds the potential to fortify the integrity of smart contracts in order to foster a more secure decentralised ecosystem, and future research in this space will only engender further trust in the power of blockchain technology.


\section{Appendix}
{
\setstretch{1}
\renewcommand{\arraystretch}{1.5}
\small 
\begin{longtable}{p{1.5cm} p{6cm}}
    \caption{Machine Learning Model Category Abbreviation Mappings} \\
    \toprule
    Abb. & Title \\
    \midrule
    \endfirsthead
    \multicolumn{2}{c}{\tablename\ \thetable{} -- Continued} \\
    \toprule
    Abb. & Title \\
    \midrule
    \endhead
    \bottomrule
    \endfoot
    \bottomrule
    \endlastfoot
    CNN & Convolutional Neural Networks \\
    GMU & General Multi-Model Usage \\
    GNN & Graph Neural Networks \\
    HYB & Hybrid Models \\
    NLP & Natural Language Processing \\
    OTH & Other Models and Techniques \\
    RNN & Recurrent Neural Networks \\
    TML & Traditional Machine Learning Techniques \\
\end{longtable}
}

{
\setstretch{1}
\renewcommand{\arraystretch}{1.5}
\small 
\begin{longtable}{p{1.5cm} p{1cm} p{6cm} p{3cm}}
    \caption{Summary of Papers} 
    \label{fig:papers_repository} \\
    \toprule
    Category & Ref. & Title & ML Models Used \\
    \midrule
    \endfirsthead
    \multicolumn{4}{c}{\tablename\ \thetable{} -- Continued} \\
    \toprule
    Category & Ref. & Title & ML Models Used \\
    \midrule
    \endhead
    \bottomrule
    \multicolumn{4}{c}{Continued on next page...} \\
    \endfoot
    \bottomrule
    \endlastfoot
    \begin{filecontents*}{csvs/papers_repository.csv}
Category,Reference,Title,ML Models Used
CNN,{{\citep{sun_attention-based_2021}}},Attention-based machine learning model for smart contract vulnerability detection,CNN
CNN,{{\citep{hwang_codenet_2022}}},CodeNet: Code-Targeted Convolutional Neural Network Architecture for Smart Contract Vulnerability Detection,CNN
CNN,{{\citep{huang_hunting_2018}}},Hunting the ethereum smart contract: Color-inspired inspection of potential attacks,CNN; DNN
CNN,{{\citep{rossini_smart_2022}}},Smart Contracts Vulnerability Classification through Deep Learning,CNN; LSTM; ResNet; SGD
CNN,{{\citep{qin_smart_2022}}},Smart Contract Vulnerability Detection Based on Critical Combination Path and Deep Learning,Lenet-5; ResNet50
CNN,{{\citep{xu_efficient_2023}}},An Efficient Code-Embedding-Based Vulnerability Detection Model for Ethereum Smart Contracts,TextCNN
CNN,{{\citep{zhou_vulnerability_2022}}},Vulnerability Analysis of Smart Contract for Blockchain-Based IoT Applications: A Machine Learning Approach,TBCNN
GMU,{{\citep{gopali_vulnerability_2022}}},Vulnerability Detection in Smart Contracts Using Deep Learning,AWD-LSTM; LSTM; TCN
GMU,{{\citep{yu_deescvhunter_2021}}},DeeSCVHunter: A Deep Learning-Based Framework for Smart Contract Vulnerability Detection,Bi-GRU; Bi-LSTM; CNN-LSTM; GRU; LSTM; LSTM-ATT; RNN; TextCNN
GMU,{{\citep{wu_detecting_2022}}},Detecting Vulnerabilities in Ethereum Smart Contracts with Deep Learning,CNN; CNN-BiLSTM; DT; KNN; LSTM; RF; ResNet
GMU,{{\citep{xing_new_2020}}},A new scheme of vulnerability analysis in smart contract with machine learning,CNN; NN; RF
GMU,{{\citep{li_smart_2022}}},Smart Contract Vulnerability Detection Based on Deep and Cross Network,DCN
GMU,{{\citep{yuan_svchecker_2022}}},SVChecker: A deep learning-based system for smart contract vulnerability detection,DL
GMU,{{\citep{deng_smart_2023}}},Smart contract vulnerability detection based on deep learning and multimodal decision fusion,DL; GCN
GMU,{{\citep{lutz_escort_2021}}},ESCORT: ethereum smart contracts vulnerability detection using deep neural network and transfer learning,DNN
GMU,{{\citep{mi_vscl_2021}}},VSCL: Automating Vulnerability Detection in Smart Contracts with Deep Learning,DNN
GMU,{{\citep{sendner_smarter_2023}}},Smarter Contracts: Detecting Vulnerabilities in Smart Contracts with Deep Transfer Learning.,DNN; GRU; LSTM
GMU,{{\citep{ashizawa_eth2vec_2021}}},Eth2Vec: Learning Contract-Wide Code Representations for Vulnerability Detection on Ethereum Smart Contracts,NN
GMU,{{\citep{xue_xfuzz_2022}}},xFuzz: Machine Learning Guided Cross-Contract Fuzzing,NB; DT; EEC; LR; LSTM; SVM; XGBoost; KNN
GMU,{{\citep{liao_soliaudit_2019}}},SoliAudit: Smart Contract Vulnerability Assessment Based on Machine Learning and Fuzz Testing,LR; SVM; KNN; DT; RF; GB; CNN
GNN,{{\citep{cai_combine_2023}}},Combine sliced joint graph with graph neural networks for smart contract vulnerability detection ,BiGGNN
GNN,{{\citep{liu_combining_2021}}},Combining Graph Neural Networks With Expert Knowledge for Smart Contract Vulnerability Detection,GNN; TMP
GNN,{{\citep{han_smart_2022}}},A smart contract vulnerability detection model based on graph neural networks,CNN; GCN; GNN
GNN,{{\citep{fan_smart_2021}}},Smart Contract Vulnerability Detection Based on Dual Attention Graph Convolutional Network,DA-GCN
GNN,{{\citep{zhuang_smart_2021}}},Smart Contract Vulnerability Detection Using Graph Neural Networks ,DR-GCN; TMP
GNN,{{\citep{nguyen_mando-guru_2022}}},MANDO-GURU: Vulnerability Detection for Smart Contract Source Code by Heterogeneous Graph Embeddings,GAT
GNN,{{\citep{luo_vuldet_2023}}},VulDet: Smart Contract Vulnerability Detection Based on Graph Attention Networks,GAT
GNN,{{\citep{zhang_toward_2022}}},Toward Vulnerability Detection for Ethereum Smart Contracts Using Graph-Matching Network,GMN; GNN
GNN,{{\citep{liu_smart_2023}}},A Smart Contract Vulnerability Detection Mechanism Based on Deep Learning and Expert Rules,GNN
GNN,{{\citep{zeng_ethergis_2022}}},EtherGIS: A Vulnerability Detection Framework for Ethereum Smart Contracts Based on Graph Learning Features,GNN
GNN,{{\citep{nguyen_mando_2022}}},Mando: Multi-level heterogeneous graph embeddings for fine-grained detection of smart contract vulnerabilities,GNN
GNN,{{\citep{wang_smart_2022}}},Smart Contract Vulnerability Detection for Educational Blockchain Based on Graph Neural Networks,GNN
GNN,{{\citep{liu_smart_2021}}},Smart contract vulnerability detection: from pure neural network to interpretable graph feature and expert pattern fusion,GNN
GNN,{{\citep{nguyen_mando-hgt_2023}}},MANDO-HGT: Heterogeneous Graph Transformers for Smart Contract Vulnerability Detection,GNN; HGT
GNN,{{\citep{han_smart_2023}}},A Smart Contract Vulnerability Detection Model Based on Syntactic and Semantic Fusion Learning,GNN; TextCNN
GNN,{{\citep{liu_vulnerable_2023}}},Vulnerable Smart Contract Function Locating Based on Multi-Relational Nested Graph Convolutional Network,MRN-GCN
GNN,{{\citep{chen_smart_2023}}},Smart contract vulnerability detection based on semantic graph and residual graph convolutional networks with edge attention,RGCN
GNN,{{\citep{wang_gvd-net_2022}}},Gvd-net: Graph embedding-based machine learning model for smart contract vulnerability detection,GNN
HYB,{{\citep{xu_vulnerability_2023}}},Vulnerability Detection of Ethereum Smart Contract Based on SolBERT-BiGRU-Attention Hybrid Neural Model,BERT-BiGRU
HYB,{{\citep{jie_novel_2023}}},A novel extended multimodal AI framework towards vulnerability detection in smart contracts,BERT; Bi-LSTM; GCN; GRU; RF; Vanilla-RNN; TextCNN
HYB,{{\citep{duy_vulnsense_2023}}},VulnSense: Efficient Vulnerability Detection in Ethereum Smart Contracts by Multimodal Learning with Graph Neural Network and Language Model,BERT; BiLSTM; GNN
HYB,{{\citep{jie_full-stack_2021}}},Full-Stack Hierarchical Fusion of Static Features for Smart Contracts Vulnerability Detection,Bi-LSTM; CNN; GCN; RF; TextCNN
HYB,{{\citep{zhang_novel_2023}}},A Novel Smart Contract Reentrancy Vulnerability Detection Model based on BiGAS,BiGRU-ATT; SVM
HYB,{{\citep{zhang_cbgru_2022}}},Cbgru: A detection method of smart contract vulnerability based on a hybrid model,BiGRU; BiLSTM; CNN; GRU; LSTM
HYB,{{\citep{wu_smart_2023}}},Smart Contract Vulnerability Detection Based on Hybrid Attention Mechanism Model,BiGRU; HAM
HYB,{{\citep{zhang_spcbig-ec_2022}}},SPCBIG-EC: a robust serial hybrid model for smart contract vulnerability detection,BiGRU; SPCNN
HYB,{{\citep{zhang_svscanner_2023}}},SVScanner: Detecting smart contract vulnerabilities via deep semantic extraction,BiLSTM; TextCNN
HYB,{{\citep{wang_detecting_2023}}},Detecting Unknown Vulnerabilities in Smart Contracts with Multi-Label Classification Model Using CNN-BiLSTM,CNN-BiLSTM
HYB,{{\citep{xu_hybrid_2023}}},A Hybrid Neural Network Model-based Approach for Detecting Smart Contract Vulnerabilities,CNN-GRU
HYB,{{\citep{li_reentrancy_2023}}},Reentrancy Vulnerability Detection Based on Conv1D-BiGRU and Expert Knowledge,Conv1DBiGRU
HYB,{{\citep{xu_w2v-sa_2023}}},W2V-SA: A Deep Neural Network-based Approach to Smart Contract Vulnerability Detection,DNN
NLP,{{\citep{sun_assbert_2023}}},ASSBert: Active and semi-supervised bert for smart contract vulnerability detection,BERT
NLP,{{\citep{jeon_smartcondetect_2021}}},Smartcondetect: Highly accurate smart contract code vulnerability detection mechanism using bert,BERT
NLP,{{\citep{duan_new_2023}}},A New Smart Contract Anomaly Detection Method by Fusing Opcode and Source Code Features for Blockchain Services,CodeBERT
NLP,{{\citep{jiang_vddl_2023}}},VDDL: A Deep Learning-Based Vulnerability Detection Model for Smart Contracts,CodeBERT
NLP,{{\citep{wu_peculiar_2021}}},Peculiar: Smart Contract Vulnerability Detection Based on Crucial Data Flow Graph and Pre-training Techniques,GraphCodeBERT
NLP,{{\citep{zhang_reentrancy_2022}}},Reentrancy Vulnerability Detection and Localization: A Deep Learning Based Two-phase Approach,GraphCodeBERT
OTH,{{\citep{narayana_automation_2023}}},Automation and smart materials in detecting smart contracts vulnerabilities in Blockchain using deep learning,ANN; Autoencoder
OTH,{{\citep{zhang_vulnerability_2022}}},Vulnerability Detection For Smart Contract Via Backward Bayesian Active Learning,BAL; TBCNN
OTH,{{\citep{zhang_novel_2022}}},A novel smart contract vulnerability detection method based on information graph and ensemble learning,Bi-GRU; CNN; DNN; Ensemble; GRU; RCNN; RNN; Transformer
OTH,{{\citep{zhang_smart_2022-1}}},Smart contract vulnerability detection combined with multi-objective detection,MODNN
OTH,{{\citep{zhou_smart_2023}}},Smart Contract Vulnerability Detection Model Based on Multi‐Task Learning,Multi-Task Learning
OTH,{{\citep{guo_smart_2022}}},Smart Contract Vulnerability Detection Model Based on Siamese Network (SCVSN): A Case Study of Reentrancy Vulnerability,SCVSN; LSTM
OTH,{{\citep{gong_gratdet_2023}}},GRATDet: Smart Contract Vulnerability Detector Based on Graph Representation and Transformer.,Transformer-GP
RNN,{{\citep{gupta_deep_2022}}},Deep learning-based malicious smart contract detection scheme for internet of things environment,ANN; GRU; LSTM
RNN,{{\citep{gogineni_multi-class_2020}}},Multi-Class classification of vulnerabilities in Smart Contracts using AWD-LSTM with pre-trained encoder inspired from natural language processing,AWD-LSTM
RNN,{{\citep{zhang_smart_2022}}},Smart Contract Vulnerability Detection Method based on Bi-LSTM Neural Network,BiLSTM
RNN,{{\citep{hu_smart-contract_2023}}},Smart-Contract Vulnerability Detection Method Based on Deep Learning,BiLSTM
RNN,{{\citep{qian_towards_2020}}},Towards Automated Reentrancy Detection for Smart Contracts Based on Sequential Models,BiLSTM-ATT
RNN,{{\citep{qian_bilstm-attention_2022}}},A BiLSTM-ATT Model for Detecting Smart Contract Defects More Accurately,BiLSTM-ATT
RNN,{{\citep{qian_multi-label_2022}}},Multi-Label Vulnerability Detection of Smart Contracts Based on Bi-LSTM and Attention Mechanism,BiLSTM-ATT
RNN,{{\citep{ren_smart_2023}}},Smart contract vulnerability detection based on a semantic code structure and a self-designed neural network,BiLSTM-ATT
RNN,{{\citep{wang_ethereum_2023}}},Ethereum Smart Contract Vulnerability Detection Model Based on Triplet Loss and BiLSTM,BiLSTM; Metric Learning
RNN,{{\citep{zhou_smart_2023}}},Smart contracts vulnerability detection model based on adversarial multi-task learning,BiLSTM; Multi-Task Learning
RNN,{{\citep{xu_reentrancy_2022}}},Reentrancy Vulnerability Detection of Smart Contract Based on Bidirectional Sequential Neural Network with Hierarchical Attention Mechanism,HAM-BiLSTM
RNN,{{\citep{goswami_tokencheck_2021}}},TokenCheck: Towards Deep Learning Based Security Vulnerability Detection In ERC-20 Tokens,LSTM
RNN,{{\citep{tann_towards_2018}}},Towards safer smart contracts: A sequence learning approach to detecting security threats,LSTM
TML,{{\citep{wang_contractward_2021}}},ContractWard: Automated Vulnerability Detection Models for Ethereum Smart Contracts,AdaBoost; KNN; RF; SVM; XGBoost
TML,{{\citep{lohith_tp-detect_2023}}},TP-Detect: trigram-pixel based vulnerability detection for Ethereum smart contracts,BRKNN; MLKNN; NB; RF
TML,{{\citep{yang_improvement_2023}}},Improvement and Optimization of Vulnerability Detection Methods for Ethernet Smart Contracts,CNN; SVM
TML,{{\citep{zhao_novel_2023}}},A Novel Machine Learning-Based Model For Reentrant Vulnerabilities Detection,DT; KNN; L; P; RDF; SVM
TML,{{\citep{momeni_machine_2019}}},Machine Learning Model for Smart Contracts Security Analysis,DT; NN; RF; SVM
TML,{{\citep{mandloi_machine_2022}}},A machine learning-based dynamic method for detecting vulnerabilities in smart contracts,DT; RF
TML,{{\citep{eshghie_dynamic_2021}}},Dynamic Vulnerability Detection on Smart Contracts Using Machine Learning,KNN; LR; NB; NN; RF; SVM
TML,{{\citep{li_detecting_2023}}},Detecting Unknown Vulnerabilities in Smart Contracts with Binary Classification Model Using Machine Learning,KNN; LR; SVM
TML,{{\citep{song_efficient_2019}}},An Efficient Vulnerability Detection Model for Ethereum Smart Contracts,KNN; RF; SVM
TML,{{\citep{lakshminarayana_towards_2022}}},Towards Auto Contract Generation and Ensemble-based Smart Contract Vulnerability Detection,KNN; RF; SVM; XGBoost
TML,{{\citep{xu_novel_2021}}},A Novel Machine Learning-Based Analysis Model for Smart Contract Vulnerability,KNN; SGD
TML,{{\citep{yang_smart_2022}}},Smart Contract Vulnerability Detection based on Abstract Syntax Tree,RF
TML,{{\citep{hao_scscan_2020}}},SCScan: A SVM-Based Scanning System for Vulnerabilities in Blockchain Smart Contracts,SVM
\end{filecontents*}
\csvreader[
        late after line=\\,
        before reading={\catcode`\_=12},
        after reading={\catcode`\_=8},
        respect underscore
    ]
    {csvs/papers_repository.csv}{}
    {\csvcoli & \csvcolii & \csvcoliii & \csvcoliv}
\end{longtable}
}

{
\setstretch{1}
\renewcommand{\arraystretch}{1.5}
\small 
\begin{longtable}{p{1cm} p{5cm} p{8cm}}
    \caption{Vulnerabilities Addressed in Each Paper} 
    \label{fig:vulnerabilities_repository} \\
    \toprule
    Ref. & Title & Vulnerabilities Addressed \\
    \midrule
    \endfirsthead
    \multicolumn{3}{c}{\tablename\ \thetable{} -- Continued} \\
    \toprule
    Ref. & Title & Vulnerabilities Addressed \\
    \midrule
    \endhead
    \bottomrule
    \multicolumn{3}{r}{Continued on next page...} \\
    \endfoot
    \bottomrule
    \endlastfoot
    \begin{filecontents*}{csvs/vulnerabilities_repository.csv}
Ref.,Title,Vulnerabilities Addressed
{{\citep{qian_bilstm-attention_2022}}},A BiLSTM-ATT Model for Detecting Smart Contract Defects More Accurately,Block Dependency; Denial Of Service; Greedy Contract; Hard Code Address; Misleading Data Location; Nested Call; Reentrancy; Strict Check For Balance; Transaction Order Dependency; Unchecked Low Level Calls; Unmatched Type Assignment
{{\citep{xu_hybrid_2023}}},A Hybrid Neural Network Model-based Approach for Detecting Smart Contract Vulnerabilities,Arithmetic; AssertFail; Block Dependency; Unchecked Low Level Calls; Reentrancy
{{\citep{mandloi_machine_2022}}},A machine learning-based dynamic method for detecting vulnerabilities in smart contracts,Reentrancy
{{\citep{duan_new_2023}}},A New Smart Contract Anomaly Detection Method by Fusing Opcode and Source Code Features for Blockchain Services,Reentrancy; Time Dependency; Transaction Order Dependency
{{\citep{jie_novel_2023}}},A novel extended multimodal AI framework towards vulnerability detection in smart contracts,Binary Classification
{{\citep{xu_novel_2021}}},A Novel Machine Learning-Based Analysis Model for Smart Contract Vulnerability,Access Control; Arithmetic; Bad Randomness; Denial Of Service; Front Running; Reentrancy; Short Address; Unchecked Low Level Calls
{{\citep{zhao_novel_2023}}},A Novel Machine Learning-Based Model For Reentrant Vulnerabilities Detection,Reentrancy
{{\citep{zhang_novel_2023}}},A Novel Smart Contract Reentrancy Vulnerability Detection Model based on BiGAS,Reentrancy
{{\citep{zhang_novel_2022}}},A novel smart contract vulnerability detection method based on information graph and ensemble learning,Callstack Depth; Arithmetic; Parity Multisig Bug 2; Reentrancy; Time Dependency; Transaction Order Dependency
{{\citep{liu_smart_2023}}},A Smart Contract Vulnerability Detection Mechanism Based on Deep Learning and Expert Rules,Gas Issues; Delegate Call; Reentrancy; Time Dependency
{{\citep{han_smart_2022}}},A smart contract vulnerability detection model based on graph neural networks,Arithmetic; ERC-20 Transfer; Implicit Visibility
{{\citep{han_smart_2023}}},A Smart Contract Vulnerability Detection Model Based on Syntactic and Semantic Fusion Learning,Arithmetic; Implicit Visibility; Reentrancy; Time Dependency
{{\citep{xu_efficient_2023}}},An Efficient Code-Embedding-Based Vulnerability Detection Model for Ethereum Smart Contracts,Arithmetic; AssertFail; Average; Callstack Depth; Checkeffects; Inline Assembly; Reentrancy; Self Destruct; Time Dependency; Transaction Order Dependency; Tx.Origin; Unchecked Low Level Calls
{{\citep{song_efficient_2019}}},An Efficient Vulnerability Detection Model for Ethereum Smart Contracts,Arithmetic; Callstack Depth; Reentrancy; Time Dependency; Transaction Order Dependency
{{\citep{sun_assbert_2023}}},ASSBert: Active and semi-supervised bert for smart contract vulnerability detection,Callstack Depth; Flow; Reentrancy; Time Dependency; Transaction Order Dependency; Tx.Origin
{{\citep{sun_attention-based_2021}}},Attention-based machine learning model for smart contract vulnerability detection,Reentrancy; Arithmetic; Timestamp Dependency
{{\citep{narayana_automation_2023}}},Automation and smart materials in detecting smart contracts vulnerabilities in Blockchain using deep learning,Denial Of Service; Reentrancy; Transaction Order Dependency
{{\citep{zhang_cbgru_2022}}},Cbgru: A detection method of smart contract vulnerability based on a hybrid model,Arithmetic; Callstack Depth; Infinite Loop; Reentrancy; Time Dependency
{{\citep{hwang_codenet_2022}}},CodeNet: Code-Targeted Convolutional Neural Network Architecture for Smart Contract Vulnerability Detection,Reentrancy; Time Dependency; Tx.Origin; Unchecked Low Level Calls
{{\citep{cai_combine_2023}}},Combine sliced joint graph with graph neural networks for smart contract vulnerability detection ,Arithmetic; Block Dependency; Delegate Call; Reentrancy; Self Destruct; Short Address; Time Dependency; Transaction Order Dependency; Unchecked Low Level Calls
{{\citep{liu_combining_2021}}},Combining Graph Neural Networks With Expert Knowledge for Smart Contract Vulnerability Detection,Infinite Loop; Reentrancy; Time Dependency
{{\citep{wang_contractward_2021}}},ContractWard: Automated Vulnerability Detection Models for Ethereum Smart Contracts,Arithmetic; Callstack Depth; Reentrancy; Time Dependency; Transaction Order Dependency
{{\citep{gupta_deep_2022}}},Deep learning-based malicious smart contract detection scheme for internet of things environment,Binary Classification
{{\citep{yu_deescvhunter_2021}}},DeeSCVHunter: A Deep Learning-Based Framework for Smart Contract Vulnerability Detection,Reentrancy; Time Dependency
{{\citep{li_detecting_2023}}},Detecting Unknown Vulnerabilities in Smart Contracts with Binary Classification Model Using Machine Learning,Block Dependency; Missing The Transfer Event; No Check After Contract Invocation; Reentrancy; Strict Check For Balance; Time Dependency; Unexpected Function Invocation
{{\citep{wu_detecting_2022}}},Detecting Vulnerabilities in Ethereum Smart Contracts with Deep Learning,Arithmetic; Callstack Depth; Reentrancy; Time Dependency; Transaction Order Dependency
{{\citep{eshghie_dynamic_2021}}},Dynamic Vulnerability Detection on Smart Contracts Using Machine Learning,Reentrancy
{{\citep{lutz_escort_2021}}},ESCORT: ethereum smart contracts vulnerability detection using deep neural network and transfer learning,AssertFail; Callstack Depth; Denial Of Service; Money Concurrency; Multiple Sends; Reentrancy; Self Destruct
{{\citep{ashizawa_eth2vec_2021}}},Eth2Vec: Learning Contract-Wide Code Representations for Vulnerability Detection on Ethereum Smart Contracts,Arithmetic; ERC-20 Transfer; Gas Issues; Implicit Visibility; Reentrancy; Time Dependency
{{\citep{wang_ethereum_2023}}},Ethereum Smart Contract Vulnerability Detection Model Based on Triplet Loss and BiLSTM,Arithmetic; Access Control; Reentrancy; Unchecked Low Level Calls
{{\citep{zeng_ethergis_2022}}},EtherGIS: A Vulnerability Detection Framework for Ethereum Smart Contracts Based on Graph Learning Features,Delegate Call; Reentrancy; Self Destruct; Time Dependency; Transaction Order Dependency; Tx.Origin
{{\citep{jie_full-stack_2021}}},Full-Stack Hierarchical Fusion of Static Features for Smart Contracts Vulnerability Detection,Binary Classification
{{\citep{gong_gratdet_2023}}},GRATDet: Smart Contract Vulnerability Detector Based on Graph Representation and Transformer.,Reentrancy
{{\citep{wang_gvd-net_2022}}},Gvd-net: Graph embedding-based machine learning model for smart contract vulnerability detection,Access Control; Arithmetic; Asset Frozen; Call To Unknown; Exception State; Gas Issues; Imutaion Bug; Random Generation; Reentrancy; Stack Size Limit; Time Dependency; Wrong State; Wrong Transfer
{{\citep{huang_hunting_2018}}},Hunting the ethereum smart contract: Color-inspired inspection of potential attacks,Compiler Version Issues; Array Access Clean Higher Order Bits; Clean Bytes Higher Order Bits; Constant Optimizer Subtraction; Delegate Call; Dynamic Allocation Infinite Loop; EC Recover Malformed Input; High Order Byte Clean Storage; Identity Precompile Return Ignored; Libraries Not Callable From Payable Functions; One Of Two Constructors Skipped; Optimizer Clear State On Code Path Join; Optimizers Stale Knowledge About Sha3; Optimizer State Knowledge Not Reset For Jumpdest; Send Fails For Zero Ether; Skip Empty String Literal; Zero Function Selector
{{\citep{yang_improvement_2023}}},Improvement and Optimization of Vulnerability Detection Methods for Ethernet Smart Contracts,Reentrancy; Unchecked Low Level Calls; Tx.Origin; Time Dependency
{{\citep{momeni_machine_2019}}},Machine Learning Model for Smart Contracts Security Analysis,Arithmetic; Inline Assembly; Locked Funds; Exception State; Incorrect ERC-20 Interface; Local Variable Shadowing; Multiple Calls In A Single Transaction; Reentrancy; Self Destruct; State Variables That Could Be Declared As Constant; Tx.Origin; Unindexed ERC-20 Event Parameters; Unchecked Low Level Calls; Usage Of Complex Pragma Statement
{{\citep{nguyen_mando-guru_2022}}},MANDO-GURU: Vulnerability Detection for Smart Contract Source Code by Heterogeneous Graph Embeddings,Access Control; Arithmetic; Denial Of Service; Front Running; Reentrancy; Time Dependency; Unchecked Low Level Calls
{{\citep{nguyen_mando-hgt_2023}}},MANDO-HGT: Heterogeneous Graph Transformers for Smart Contract Vulnerability Detection,Access Control; Arithmetic; Denial Of Service; Front Running; Reentrancy; Front Running; Time Dependency; Unchecked Low Level Calls
{{\citep{nguyen_mando_2022}}},Mando: Multi-level heterogeneous graph embeddings for fine-grained detection of smart contract vulnerabilities,Access Control; Arithmetic; Denial Of Service; Front Running; Reentrancy; Time Dependency; Unchecked Low Level Calls
{{\citep{gogineni_multi-class_2020}}},Multi-Class classification of vulnerabilities in Smart Contracts using AWD-LSTM with pre-trained encoder inspired from natural language processing,SPGN Classification
{{\citep{qian_multi-label_2022}}},Multi-Label Vulnerability Detection of Smart Contracts Based on BiLSTM and Attention Mechanism,Arithmetic; Reentrancy; Time Dependency; Transaction Order Dependency
{{\citep{wu_peculiar_2021}}},Peculiar: Smart Contract Vulnerability Detection Based on Crucial Data Flow Graph and Pre-training Techniques,Reentrancy
{{\citep{zhang_reentrancy_2022}}},Reentrancy Vulnerability Detection and Localization: A Deep Learning Based Two-phase Approach,Reentrancy
{{\citep{li_reentrancy_2023}}},Reentrancy Vulnerability Detection Based on Conv1D-BiGRU and Expert Knowledge,Reentrancy
{{\citep{xu_reentrancy_2022}}},Reentrancy Vulnerability Detection of Smart Contract Based on Bidirectional Sequential Neural Network with Hierarchical Attention Mechanism,Reentrancy
{{\citep{hao_scscan_2020}}},SCScan: A SVM-Based Scanning System for Vulnerabilities in Blockchain Smart Contracts,Access Control; Arithmetic; Denial Of Service; Predictable Random Number Attack; Reentrancy; Unchecked Low Level Calls
{{\citep{ren_smart_2023}}},Smart contract vulnerability detection based on a semantic code structure and a self-designed neural network,Arithmetic; Delegate Call; Reentrancy; Time Dependency
{{\citep{yang_smart_2022}}},Smart Contract Vulnerability Detection based on Abstract Syntax Tree,Bad Randomness; Reentrancy; Unchecked Low Level Calls
{{\citep{qin_smart_2022}}},Smart Contract Vulnerability Detection Based on Critical Combination Path and Deep Learning,Arithmetic; Reentrancy; Time Dependency; Unrestricted Operational Vulnerability
{{\citep{li_smart_2022}}},Smart Contract Vulnerability Detection Based on Deep and Cross Network,Infinite Loop; Reentrancy; Time Dependency
{{\citep{deng_smart_2023}}},Smart contract vulnerability detection based on deep learning and multimodal decision fusion,Arithmetic; Locked Funds; Reentrancy; Transaction Order Dependency
{{\citep{fan_smart_2021}}},Smart Contract Vulnerability Detection Based on Dual Attention Graph Convolutional Network,Reentrancy; Time Dependency
{{\citep{wu_smart_2023}}},Smart Contract Vulnerability Detection Based on Hybrid Attention Mechanism Model,Binary Classification; Arithmetic; Reentrancy; Time Dependency; Unchecked Low Level Calls; Tx.Origin
{{\citep{chen_smart_2023}}},Smart contract vulnerability detection based on semantic graph and residual graph convolutional networks with edge attention,Arithmetic; Reentrancy; Time Dependency; Unchecked Low Level Calls
{{\citep{zhang_smart_2022-1}}},Smart contract vulnerability detection combined with multi-objective detection,Arithmetic; Checkeffects; Inline Assembly; Block Dependency; Unchecked Low Level Calls; Time Dependency; Transaction Order Dependency
{{\citep{wang_smart_2022}}},Smart Contract Vulnerability Detection for Educational Blockchain Based on Graph Neural Networks,Binary Classification
{{\citep{zhang_smart_2022}}},Smart Contract Vulnerability Detection Method based on BiLSTM Neural Network,Arithmetic; Delegate Call; Reentrancy; Time Dependency
{{\citep{zhou_smart_2023}}},Smart Contract Vulnerability Detection Model Based on Multi‐Task Learning,Arithmetic; Unknown Address Vulnerability; Reentrancy
{{\citep{guo_smart_2022}}},Smart Contract Vulnerability Detection Model Based on Siamese Network (SCVSN): A Case Study of Reentrancy Vulnerability,Reentrancy
{{\citep{zhuang_smart_2021}}},Smart Contract Vulnerability Detection Using Graph Neural Networks ,Infinite Loop; Reentrancy; Time Dependency
{{\citep{liu_smart_2021}}},Smart contract vulnerability detection: from pure neural network to interpretable graph feature and expert pattern fusion,Infinite Loop; Reentrancy; Time Dependency
{{\citep{rossini_smart_2022}}},Smart Contracts Vulnerability Classification through Deep Learning,Access Control; Arithmetic; Other; Reentrancy; Unchecked Low Level Calls
{{\citep{zhou_smart_2023}}},Smart contracts vulnerability detection model based on adversarial multi-task learning,Arithmetic; Reentrancy; Unknown Address Vulnerability
{{\citep{hu_smart-contract_2023}}},Smart-Contract Vulnerability Detection Method Based on Deep Learning,Arithmetic; Reentrancy; Time Dependency
{{\citep{jeon_smartcondetect_2021}}},Smartcondetect: Highly accurate smart contract code vulnerability detection mechanism using bert,Approve Function In ERC-20 Library; Array Length Manipulation; Attacker Can Make ERC-20 Functions Always Return False; Deprecated Constructions; Hard Code Address; Function That Have Return Value Should Not Be Internal And Private; Function That Should Not Be Pure; Function That Should Not Be Viewable; Functions Returns Incorrect Type Or Nothing; Gas Issues; Locked Funds; Msg.Value Equals Zero; Private Doesn't Hide Data; Reentrancy; Revert Require; Safemath Library (Sml); Strict Check For Balance; Compiler Version Issues; Transfer Function In Loop (TFL); Unchecked Low Level Calls; Inline Assembly; Var; Implicit Visibility
{{\citep{sendner_smarter_2023}}},Smarter Contracts: Detecting Vulnerabilities in Smart Contracts with Deep Transfer Learning.,Arbitrary Jump With Function Type Variable; Arithmetic; AssertFail; Callstack Depth; Denial Of Service; Money Concurrency; Reentrancy; Self Destruct; Time Dependency; Unchecked Low Level Calls; Bad Randomness
{{\citep{liao_soliaudit_2019}}},SoliAudit: Smart Contract Vulnerability Assessment Based on Machine Learning and Fuzz Testing,AssertFail; Block Dependency; Callstack Depth; Checkeffects; Inline Assembly; Unchecked Low Level Calls; Reentrancy; Self Destruct; Time Dependency; Transaction Order Dependency; Tx.Origin; Arithmetic
{{\citep{zhang_spcbig-ec_2022}}},SPCBIG-EC: a robust serial hybrid model for smart contract vulnerability detection,Arithmetic; Callstack Depth; Infinite Loop; Reentrancy; Time Dependency
{{\citep{yuan_svchecker_2022}}},SVChecker: A deep learning-based system for smart contract vulnerability detection,Access Control; Arithmetic; Front Running; Unchecked Low Level Calls; Reentrancy; Time Dependency
{{\citep{zhang_svscanner_2023}}},SVScanner: Detecting smart contract vulnerabilities via deep semantic extraction,Binary Classification
{{\citep{goswami_tokencheck_2021}}},TokenCheck: Towards Deep Learning Based Security Vulnerability Detection In ERC-20 Tokens,SPGN Classification
{{\citep{zhang_toward_2022}}},Toward Vulnerability Detection for Ethereum Smart Contracts Using Graph-Matching Network,Block Dependency; Reentrancy; Time Dependency; Tx.Origin
{{\citep{lakshminarayana_towards_2022}}},Towards Auto Contract Generation and Ensemble-based Smart Contract Vulnerability Detection,Denial Of Service; Reentrancy; Tx.Origin; Unchecked Low Level Calls
{{\citep{qian_towards_2020}}},Towards Automated Reentrancy Detection for Smart Contracts Based on Sequential Models,Reentrancy
{{\citep{tann_towards_2018}}},Towards safer smart contracts: A sequence learning approach to detecting security threats,SPGN Classification
{{\citep{lohith_tp-detect_2023}}},TP-Detect: trigram-pixel based vulnerability detection for Ethereum smart contracts,Arithmetic; Callstack Depth; Parity Multisig Bug 2; Reentrancy; Time Dependency; Transaction Order Dependency
{{\citep{jiang_vddl_2023}}},VDDL: A Deep Learning-Based Vulnerability Detection Model for Smart Contracts,Strict Check For Balance; Byte Array; Compiler Version Issues; Costly Loop; Denial Of Service; Implicit Visibility; Arithmetic; Locked Funds; Malicious Libraries; Private Modifier; Reentrancy; Send Instead Of Transfer; Style Guide Violation; Time Dependency; Token Api Violation; Gas Issues; Tx.Origin; Unchecked Low Level Calls; Unsafe Type Inference
{{\citep{mi_vscl_2021}}},VSCL: Automating Vulnerability Detection in Smart Contracts with Deep Learning,Binary Classification
{{\citep{luo_vuldet_2023}}},VulDet: Smart Contract Vulnerability Detection Based on Graph Attention Networks,Reentrancy; Time Dependency
{{\citep{zhou_vulnerability_2022}}},Vulnerability Analysis of Smart Contract for Blockchain-Based IoT Applications: A Machine Learning Approach,Arithmetic; Exception State; Reentrancy; Bad Randomness
{{\citep{zhang_vulnerability_2022}}},Vulnerability Detection For Smart Contract Via Backward Bayesian Active Learning,Block Dependency; Callstack Depth; Arithmetic; Reentrancy; Time Dependency; Transaction Order Dependency; Tx.Origin
{{\citep{gopali_vulnerability_2022}}},Vulnerability Detection in Smart Contracts Using Deep Learning,SPGN Classification
{{\citep{xu_vulnerability_2023}}},Vulnerability Detection of Ethereum Smart Contract Based on SolBERT-BiGRU-Attention Hybrid Neural Model,Reentrancy; Time Dependency; Transaction Order Dependency; Unchecked Low Level Calls; Arithmetic
{{\citep{liu_vulnerable_2023}}},Vulnerable Smart Contract Function Locating Based on Multi-Relational Nested Graph Convolutional Network,Arithmetic; Reentrancy; Time Dependency
{{\citep{duy_vulnsense_2023}}},VulnSense: Efficient Vulnerability Detection in Ethereum Smart Contracts by Multimodal Learning with Graph Neural Network and Language Model,Arithmetic; Reentrancy
{{\citep{xu_w2v-sa_2023}}},W2V-SA: A Deep Neural Network-based Approach to Smart Contract Vulnerability Detection,Arithmetic; AssertFail; Block Dependency; Checkeffects; Unchecked Low Level Calls; Reentrancy
{{\citep{xue_xfuzz_2022}}},xFuzz: Machine Learning Guided Cross-Contract Fuzzing,Delegate Call; Reentrancy; Tx.Origin
\end{filecontents*}
\csvreader[
        late after line=\\,
        before reading={\catcode`\_=12},
        after reading={\catcode`\_=8},
        respect underscore
    ]
    {csvs/vulnerabilities_repository.csv}{}
    {\csvcoli & \csvcolii & \csvcoliii}
\end{longtable}
}

\bibliographystyle{abbrvnat}
\bibliography{Main}

\end{document}